\documentclass[prb,twocolumn,aps,longbibliography]{revtex4-2}
\usepackage{graphicx,amsmath,xcolor}
\usepackage{subcaption}
\usepackage{hyperref}
\usepackage{float}
\usepackage{booktabs}
\usepackage{multirow}
\usepackage{mwe}
\usepackage{silence}
\usepackage{mathtools}
\usepackage[T1]{fontenc}

\begin{document}

    \title{Theory of charge stability diagrams in coupled quantum dot qubits}
    \author{Nathan L. Foulk}
    \author{Sankar Das Sarma}
    \affiliation{Condensed Matter Theory Center and Joint Quantum Institute, Department of Physics, University of Maryland, College Park, Maryland 20742-4111 USA}
\begin{abstract}
 We predict large regions of the charge stability diagram using a multi-band and multi-electron configuration interaction model of a double quantum dot system.
 We account for many-body interactions within each quantum dot using full configuration interaction and solve for single-particle density operators.
 This allows charge states to be predicted more accurately than the extensively used classical capacitance model or the single-band Hubbard model. 
 The resulting single-particle mixed states then serve as inputs into an atomic orbital picture that allows for the explicit calculation of the underlying Hubbard model parameters by performing the appropriate integrals. This numerical approach allows for arbitrary choices of electrostatic potential and gate geometry.
 A common assumption when calculating charge stability diagrams from the Hubbard model is that the charge stability diagrams are periodic, but we find that the tunnel couplings for valence electrons in dots with $N=3$ electrons are significantly enhanced when compared to single-electron dots. 
 This difference is apparent in the charge stability diagram for higher occupancy Coulomb diamonds. 
 We also quantitatively explore how the barrier gate strength and dot pitch impact this behavior.
 Our work should help improve the future realistic modeling of semiconductor-dot-based quantum circuits.
\end{abstract}
\maketitle 
\section{Introduction}
Quantum dot based semiconductor spin qubits~\cite{hanson_spins_2007,burkard_semiconductor_2023} are a promising candidate for scalable quantum computing, benefitting from long coherence times~\cite{hansen_implementation_2022,veldhorst_addressable_2014,mills_high-fidelity_2022}, fast gate operations ~\cite{takeda_fault-tolerant_2016,mills_two-qubit_2022,croot_flopping-mode_2020,weinstein_universal_2023}, and integration possibilities with the existing semiconductor industry~\cite{neyens_probing_2024, elsayed_low_2022,ha_flexible_2022}.
Using lithographically defined metal gates, quantum dots are formed in the 2D quantum well of a semiconductor heterostructure.
These dots are extremely small--on the order of 100 nm--and their small size provides much of the promise as one of the few qubit implementations that are serious candidates for a fully error-corrected and scalable quantum computer~\cite{fowler_surface_2012,vandersypen_interfacing_2017}---the scalability because of the compatibility with the vast semiconductor electronics industry.
Each dot is loaded with a predetermined number of electrons, and the qubits are represented by the spin states of these electrons.
Single spin manipulation can be performed with electrical or magnetic means and the Heisenberg exchange interaction allows for fast and predictable electrical two-spin control between nearest neighbors~\cite{koppens_spin_2008,golovach_electric-dipole-induced_2006, petta_coherent_2005}.

\begin{figure*}
  \centering
  \begin{tabular}[c]{cc}
    \multirow{2}{*}[35mm]{
      \includegraphics[width=0.5
      \textwidth]{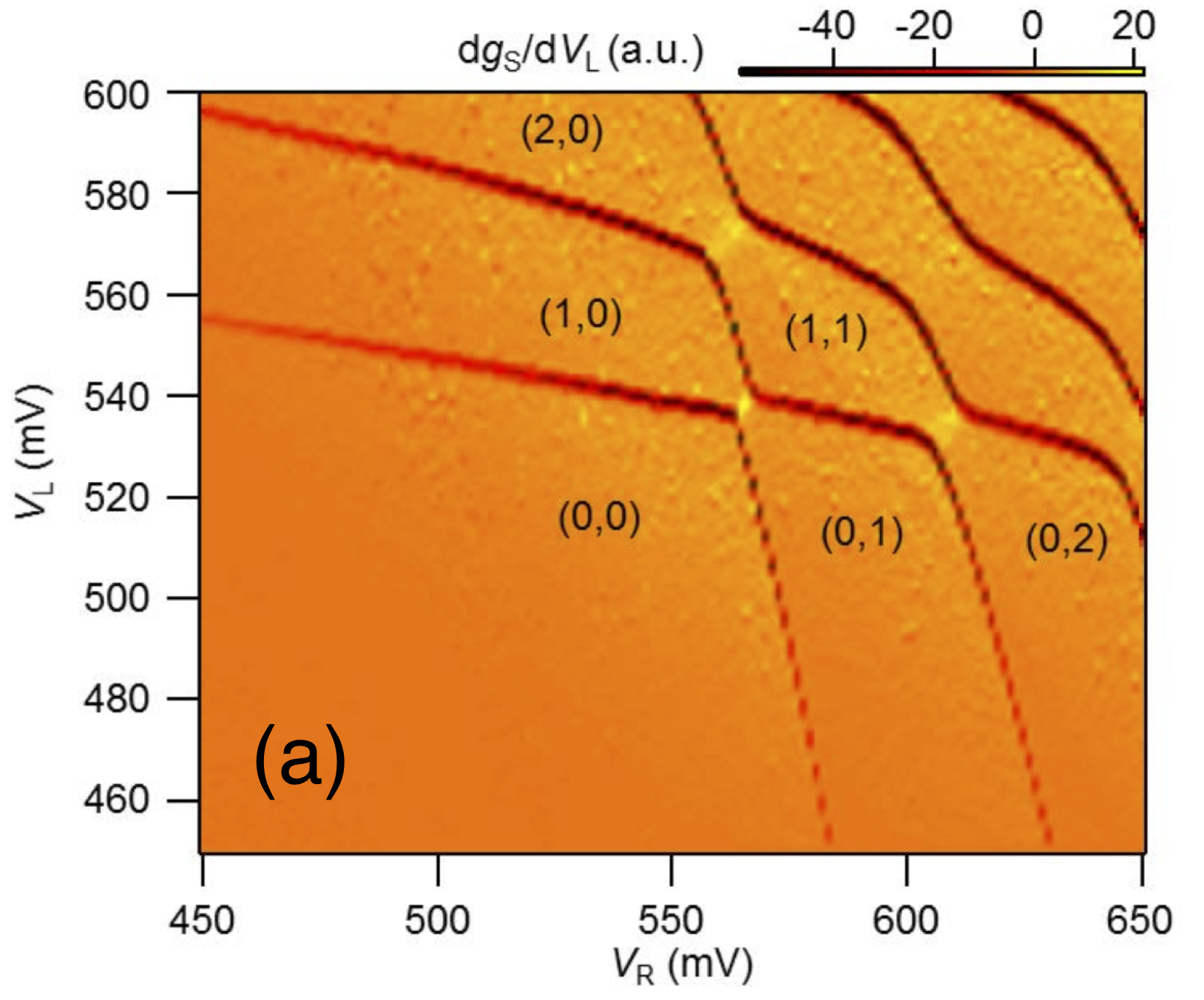}
 } &
      \includegraphics[width=0.42\textwidth]{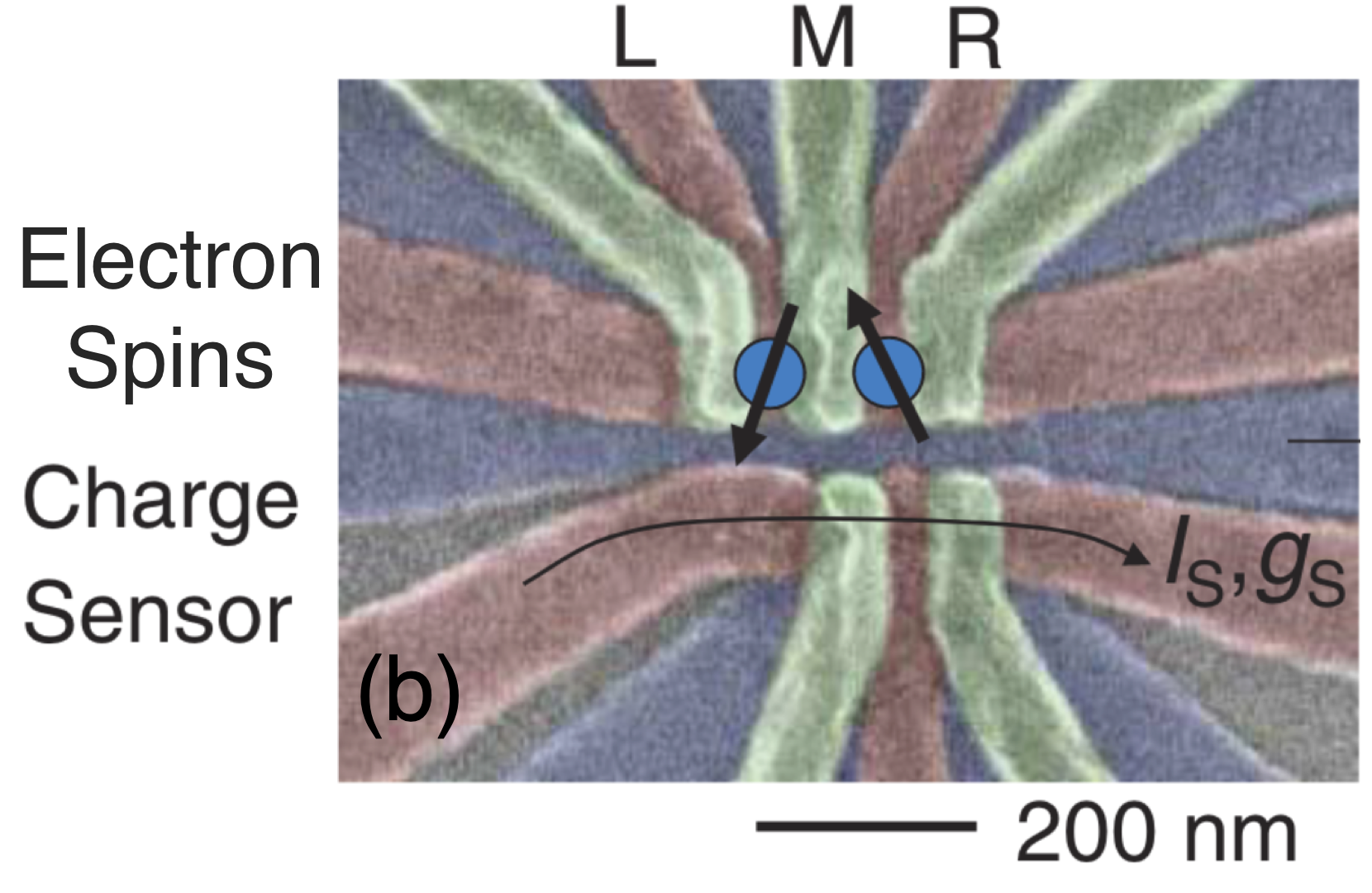}
    \\
    &
      \;\!\!\includegraphics[width=0.4\textwidth]{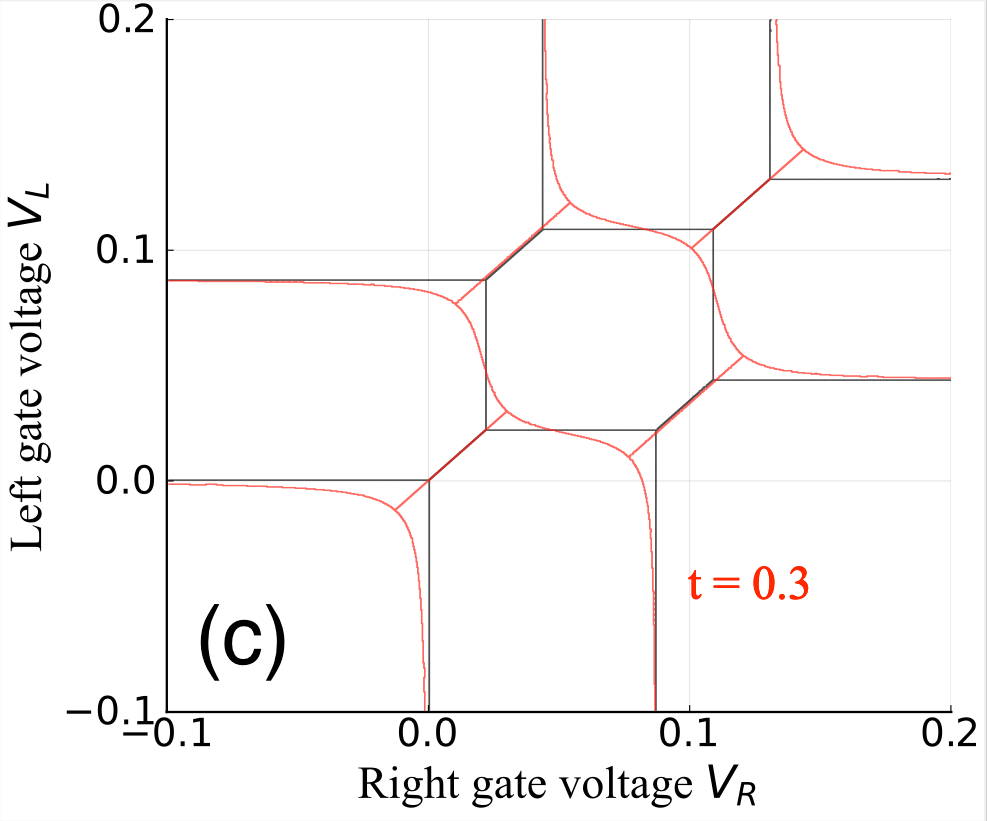}
    \\
  \end{tabular}    
  \caption{\textbf{(a)} An example of an experimental charge stability diagram. Dark red portions indicate transitions of the total system's charge. For the first few Coulomb diamonds, yellow transitions are also visible which demarcate dot-level charge transitions. \textbf{(b)} The accompanying double quantum dot device. $L$ and $R$ label to the left and right plunger gates, respectively. $M$ labels the barrier gate between them. The dot at the bottom is the charge sensor, which allows for the direct measurement of charge stability diagrams.
  The electron spins, which store quantum information, reside in the Si well under the plunger gates. Pictorial representations of the electrons and their spin projections are drawn in the form of blue dots and vectors to indicate their approximate locations within this device.
  Device imaging and charge stability data from Zajac et al. (2018)~\cite{zajac_resonantly_2018} \textbf{(c)} A theoretical charge stability diagram using the Hubbard model with $U_i = 2$ meV, $V=0.5$ meV, and $t = 0.0$ meV (black), $0.3$ meV (red). Increased tunnel coupling leads to greater smoothing of the corners of the Coulomb diamond.}
  \label{fig:intro_fig}
\end{figure*}

The simplest qubit encoding is the Loss-DiVincenzo qubit, where the two-level system is a single electron spin confined to a single quantum dot~\cite{loss_quantum_1998,hu_hilbert-space_2000,burkard_coupled_1999}.
Many other qubit encodings exist, including singlet-triplet qubits~\cite{petta_coherent_2005}, exchange-only qubits~\cite{medford_self-consistent_2013}, and the flopping-mode qubit~\cite{benito_electric-field_2019} to name a few.
However, all encoding schemes require precise control over the electron number of each dot (although a restriction to one electron per dot is not essential for qubit operations) \cite{hu_spin-based_2001,nielsen_six-electron_2013}.
Therefore, one must tune each device so that each dot contains the correct electron number, and this must be done each time the device is cooled and initialized.
In addition, this number of electrons must be preserved during the gate operations---a 1-electron dot cannot suddenly become a 3- or 5-electron dot due to local fluctuations.
A typical spin qubit device contains multiple dots arranged in a linear array using an alternating pattern of plunger gates and barrier gates.
An example of such a device is found in Fig.~\ref{fig:intro_fig}(b).
Each quantum dot is electrostatically defined under a plunger gate.
Plunger gate voltages are raised or lowered to control the electron number of each dot.
The two dots are separated by a distance known as the pitch.
Barrier gates (between the dots) are primarily used to modulate the wavefunction overlap between two dots.
The gate voltages on the plunger and barrier gates are the most important variables in determining the charge state of the dots.
Modern quantum dot devices are made of many more gates and components than just these two, but the plunger and barrier gates occupy a central role.
Any minimal model must include the role of these two gates, one controlling each dot and the other controlling the inter-dot coupling.

Charge stability diagrams are used to visualize and map out the different charge regimes in a double quantum dot, thus enabling the precise control over electron occupancy necessary for qubit operations.
The plunger gate voltages for two neighboring dots are varied, and the conductance is measured. 
A nonzero conductance signifies that electrons are tunneling into the device or between dots. 
Modern devices also employ nearby quantum dots for charge-sensing measurements, where the charge-sensor signal measures the charge states capacitively.
An example of a charge stability diagram is shown in Fig.~\ref{fig:intro_fig}a.
The diagram typically consists of sharp peaks of transitions and large plateaus of charge state stability.
These different plateaus demarcate different charge states for a double dot system. Each plateau, or cell---referred to as a ``Coulomb diamond.''---indicates fixed electron numbers.

These charge stability diagrams are essential for determining not only the initial gate voltages necessary for reliable charge initialization but also for mapping paths across the stability diagram, which are needed to carry out precise gate operations on the qubits.
Recently, charge stability diagrams have also been used as inputs to machine learning approaches for automated control of quantum dot arrays~\cite{taylor_neural_2024,ziegler_tuning_2023,zwolak_colloquium_2023}.
The direct calculation of realistic charge stability diagrams given the underlying microscopics of a device is therefore a timely need in the quantum computing research community in order to cost-effectively produce qubit designs and to generate input data for machine learning algorithms for autonomous control.

As a complement to direct measurement, one can predict the charge boundaries of the stability diagram. 
This can be done on varying levels of sophistication, the simplest of which is the classical capacitance model of the double dot system~\cite{hanson_spins_2007}.
The classical capacitance model uses the physical dimensions of the dots and their pitch to estimate the electrostatic parameters of the system, such as the charging energy of each dot and the Coulomb interaction between the dots.
This approach correctly identifies the charge regimes in the absence of quantum fluctuations. 
In order to include the effects of quantum fluctuations---which are obviously significant for double quantum dot systems---one can use the Hubbard model with nonzero tunneling~\cite{yang_generic_2011}.
The major Hubbard parameters that affect the stability diagram are the onsite Coulomb repulsion $U$, the interdot Coulomb repulsion $V$, and the tunnel coupling $t$.
The generalized Hubbard model reproduces the capacitance model exactly when $t = 0$. 
When $t > 0$, however, the sharp corners of the diamonds become rounded, with larger tunnel constants leading to more rounding (see Fig.~\ref{fig:intro_fig}c).
The principal difficulty remains in choosing realistic Hubbard parameters for a given physical system. 
Thus, the generalized Hubbard model, introduced in this context in Refs.~\cite{yang_generic_2011,wang_quantum_2011,das_sarma_hubbard_2011}, considerably simplifies charge stability modeling provided that the microscopic parameters $t$, $U$, and $V$ are known for the quantum dot system.

In the simplest case, given model confinement potentials and single electron occupancy, the estimation of these effective Hubbard parameters reduces to integrals of the single particle eigenstates~\cite{yang_generic_2011,das_sarma_hubbard_2011,wang_quantum_2011}.
Such an approach is also implicitly based on charge stability diagram periodicity.
One chooses a single point $(v_L,v_R)$ within the $(N_L,N_R) = (1,1)$ cell, calculates the wavefunctions of each dot separately, and estimates the corresponding Hubbard parameters, consistent with a Hund-Mulliken atomic orbital picture.
Those parameters are then extended to the entire stability diagram in a way that produces perfectly periodic Coulomb diamonds.
Experimental stability diagrams superficially justify this approach, as their Coulomb diamonds are often close to periodic with regard to cell size and location. 
However, the smoothness of the cell corners often varies drastically across the diagram~\cite{zajac_resonantly_2018,ziegler_tuning_2023, zwolak_autotuning_2020}.
In order to improve this approach of charge stability diagram simulation, calculations featuring multiple electrons in each dot must be included so as to go beyond the $(N_L,N_R) = (1,1)$ Coulomb diamond constraint.
We would expect the Hubbard parameters to depend on device-specific details, such as gate geometry, barrier gate level, electron occupancy, and both magnetic and electrostatic disorder. 
Incorporating such device-specific details requires the use of extensive, system-dependent numerics, rather than analytic approaches.

Going beyond the $(N_L,N_R) = (1,1)$ Coulomb diamond also addresses the practical needs of the spin qubit community, since a common practice is to load a quantum dot with three electrons rather than one (in principle, any odd number of electrons per dot suffices, but three seems to be the optimal number).
The spin state of the valence electron is then used as the qubit's two-level system~\cite{hu_spin-based_2001,barnes_screening_2011}.
Calculating the Hubbard parameters of multi-electron dots is significantly more complex than single-electron dots. 
The presence of other electrons perturbs the valence electron's wavefunction from the simple analytic expressions often used when assuming simplified potentials.
This increased electron number also complicates numerical treatments of the system, for the simple fact that simulating six-electron systems is very computationally expensive, as the relevant Hilbert space grows exponentially with the number of electrons.
An approach that leverages the computational efficiency of the atomic orbital projection method of previous work toward understanding $N=3$ quantum dots is therefore needed. 
This is our goal in the current work utilizing the configuration interaction method, which is the standard technique in quantum chemistry for calculating the electronic structure of multielectron atoms and molecules.

We first present our approach of using full configuration interaction (FCI) to project an accurate representation of $N=3$ quantum dots in the effective Hubbard model space for simulating charge stability diagrams.
Separating the double dot system into two effective single dot potentials, we solve for the three electron wavefunction of each dot.
Using the natural orbitals of the one-electron reduced density matrix (1RDM) of each dot, we approximate the atomic orbitals and calculate the relevant Hubbard parameters.
We then present our results, emphasizing the novelties introduced by the many-body treatment, including enhanced valence tunnel coupling and an understanding of how barrier gate strength and device layout impact the system's projection onto the Hubbard model.
Finally, we discuss the implications of our findings for simulating charge stability diagrams and the quantum dot semiconductor spin qubits field at large.
Our explicit results assume the qubit platform to be based on quantum dots on the Si [100] surface, but the theory and the general method are applicable to all semiconductor quantum dot structures.

\section{Model}

We calculate charge stability diagrams of double quantum dot systems using a generalized Hubbard model.

\begin{multline}
 \hat{H} =  \mu_1 n_{1} +  \mu_2 n_{2} + \sum_i\frac{U}{2} n_i (n_i - 1) + V n_1 n_2  \\ 
 + \sum_m \left(t_{m} c^\dagger_{1m} c^{}_{2m} + \text{H.c.}\right)
 \label{eq:bigham}
\end{multline}
where $n_i $ is the total electron dot occupancy over all levels for dot $i \in \{1,2\}$, and $c^\dagger_{ij}$ is the creation operator for the $j$th level for the $i$th dot. 
Many terms can be included in a generalized Hubbard model, but the terms that dominate the charge stability diagram are $t$, $U$, and $V$, which are necessary in the minimal effective model (and should suffice in most situations) since interaction terms beyond nearest-neighbors are likely to be vanishingly small.
All other terms can be completely ignored, and the stability diagram will be almost exactly the same.
This Hamiltonian can be rewritten as 
\begin{equation}
 \hat{H} = \sum_{i,j} F_{ij} c_i^\dagger c_j + \sum_{i,j,k,l} G_{ijkl} c_i^\dagger c^\dagger_j c_k c_l. 
\end{equation}
Given the appropriate single particle wavefunctions that correspond to $c_i^\dagger|0\rangle$ and $c_j^\dagger|0\rangle$, where $|0\rangle$ is the vacuum state, we can calculate $F_{ij}$
\begin{equation}
 F_{ij} = \int \Psi_i^*(\mathbf{r}) \hat{H} \Psi_j(\mathbf{r}) d \mathbf{r}.
\end{equation}
This can easily be extended to two-body terms as well: 
\begin{equation}
 G_{ijkl} = \iint \Psi_i^*(\mathbf{r})\Psi_j^*(\mathbf{r'}) \hat{H} \Psi_k(\mathbf{r}) \Psi_l(\mathbf{r'}) d\mathbf{r} d\mathbf{r}'.
\end{equation}
However, the many-body system is more complex than a product state of single-particle wavefunctions, and care must be taken to extract realistic single-particle wavefunctions that are mutually orthogonal to each other from the many-body state.
We first calculate the many-body wavefunction.

We utilize full configuration interaction (FCI)~\cite{szabo_modern_1982} to calculate the multi-electron ground state. 
These FCI calculations are performed within the effective mass approximation using the following Hamiltonian: 
\begin{multline}
 H(\mathbf{r}) = \sum_i \left(\frac{\left(\mathbf{p}_i^2 + e\mathbf{A}(\mathbf{r}_i)\right)^2}{2m_e^*}  + V(\mathbf{r}_i) + \mu_B \mathbf{S}_i \cdot \mathbf{B} \right)\\
 + \sum_{i,j} \frac{e^2}{4\pi\epsilon}\frac{1}{|\mathbf{r}_i - \mathbf{r}_j|} .
 \label{eq:fci}
\end{multline}
where $\mathbf{S}_i$ is the spin angular momentum operator of each electron, and $m^*_e$ is the effective mass for silicon in the in-plane directions, approximately $m^*_e \approx 0.19m_0$, where $m_0$ is the electron's rest mass. 
Additionally, electrostatic interactions are modulated by the increased permittivity for silicon, which is $\epsilon \approx 11.68 \epsilon_0$.
The Zeeman energy term has an insignificant effect on the charge stability diagram for realistic values of magnetic field strength. 
Most spin qubit architectures however, do operate with a finite magnetic field, and its presence simplifies our numerics by splitting spin degeneracies with a modest Zeeman splitting.
We utilize an analytic expression for the electrostatic potential in a quantum well from a square gate of radius $a$ at a distance $z$ between the gate and the well~\cite{davies_modeling_1995,anderson_high-precision_2022}

\begin{figure}
  \includegraphics*[width=0.95\linewidth]{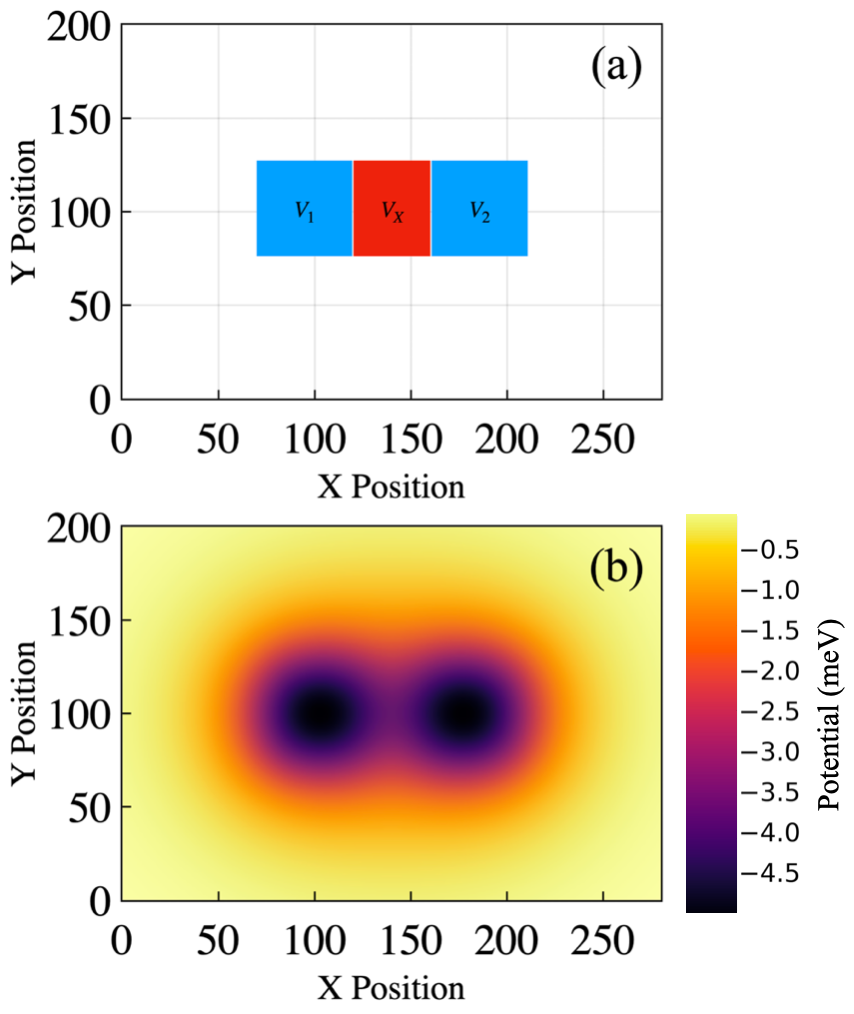}
  \caption{(a) A schematic of an example double quantum dot gate architecture. The surrounding $200\times280$ nm is the space on which the Fourier basis functions are defined.
  (b) A heatmap of the potential landscape $V_\text{total}(x,y)$ with $v_1 = v_2 = 0.2$ V and $v_X = 0$ V. 
  $X$ and $Y$ positions are in units of nanometers.}
  \label{fig:schematic}
\end{figure}

\begin{equation}
\begin{aligned}
  V_i(x,y) = v_i \bigl[g&(x-a, y-a,z) \\
  &+ g(x-a, a -y,z) \\
  &+ g(a-x, y-a,z) \\
  &+ g(a - x, a -y,z)\bigr]\biggr|_{z = 30\text{ nm}}
  \label{eq:dotPotential}
\end{aligned}
\end{equation}
where
\begin{equation}
 g(x,y,z) = \frac{1}{2\pi}\arctan \frac{xy}{z\sqrt{x^2 + y^2 + z^2}},
\end{equation}
$v_i$ is the voltage applied to the gate electrode, and $i \in \{1,2\}$ corresponds to the dot label. The coordinate $z$ is dropped from the arguments of $V_i$ since we set the $z$ coordinate to $30$ nm for our 2D calculations. An example of a possible gate architecture with its accompanying electrostatic potential is shown in Fig.~\ref{fig:schematic}.
The $30$ nm gap between the gate electrodes themselves and the quantum well, combined with the increased permittivity of silicon leads to an electrostatic potential energy $V(x,y)$ on the order of several meV.
For simplicity of demonstration, we will focus on square gates. 
For all of our calculations, each plunger gate is $50\times50$ nm, and the barrier gate is $30\times50$ nm as shown in Fig.~\ref{fig:schematic}a.
We emphasize that these choices are typical for the currently experimentally used Si-based quantum dot qubits, and our qualitative results are independent of these specific numerical choices for the confining potential. 
Our approach can be implemented using any electrostatic potential.
Gates of different shapes, sizes, and pitches can be used, the only thing needed is a grid of the confining potential producing the dots.
This flexibility makes the potential incorporation of electrostatic disorder in the calculation of Hubbard parameters and charge stability diagrams relatively straightforward, though a study of electrostatic disorder is beyond the scope of this work.
Our choice of potential is arbitrary and could use any level of sophistication. 
The potential used in this technique could easily be solved by carrying out the standard Schrodinger-Poisson self-consistent calculation using the relevant lithographic geometry. 
However, we use the model potential confinement above for the dots instead of doing a numerical Schrodinger-Poisson approach because we are presenting a general theory instead of modeling specific devices.
We note as an aside that the Schrodinger-Poisson technique is essentially a one-particle problem, and is not particularly computationally demanding compared with the FCI technique we employ for extracting the Hubbard parameters as discussed below.

In the case of multi-electron dots, calculating the exact many-body wavefunction for the double quantum dot becomes prohibitively expensive (even for our model confinement potential). 
This is because the minimum number for each dot is three in order to have a valence electron, and $N=6$ between the two dots is expensive enough to make an FCI calculation impractical. 
This leads us instead to treat the intradot and interdot interactions separately. 
We assume that the correlation effects between the two dots are small enough to be safely ignored for the FCI calculations, and we calculate the FCI ground state for each dot separately. This is also consistent with the atomic orbital approach for calculating the Hubbard parameters since the generalized Hubbard approach to spin qubits implicitly assumes this individual dot independence we are using in our FCI theory.

\begin{figure*}
    \begin{subfigure}{.5\textwidth}
      \centering
      \includegraphics[width=.9\linewidth]{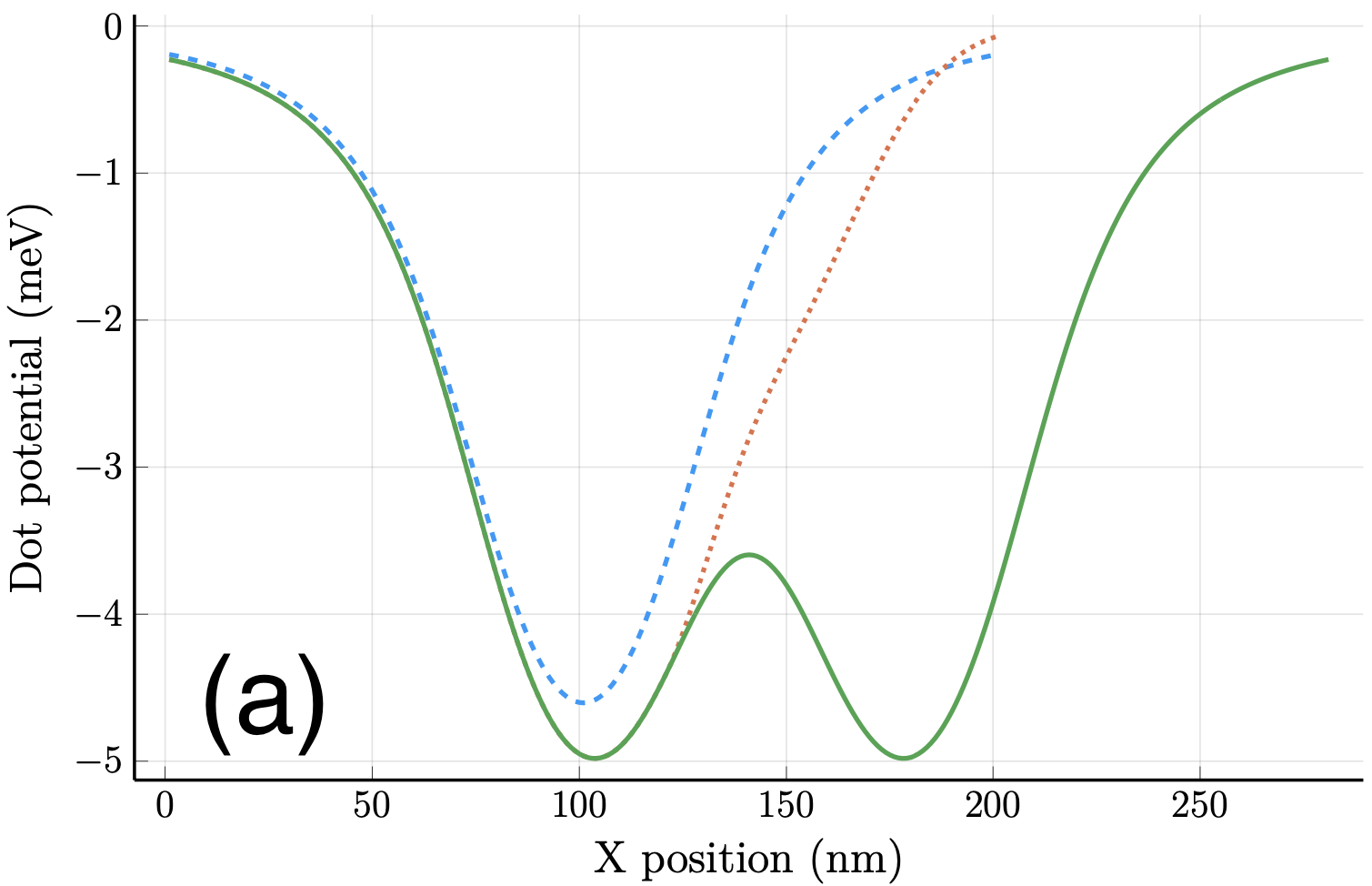}
      \label{fig:sfig1}
      \vspace{5mm}
    \end{subfigure}%
    \!\!\!\!\!\!\!\!
    \begin{subfigure}{.5\textwidth}
      \centering
      \includegraphics[width=1.03\linewidth]{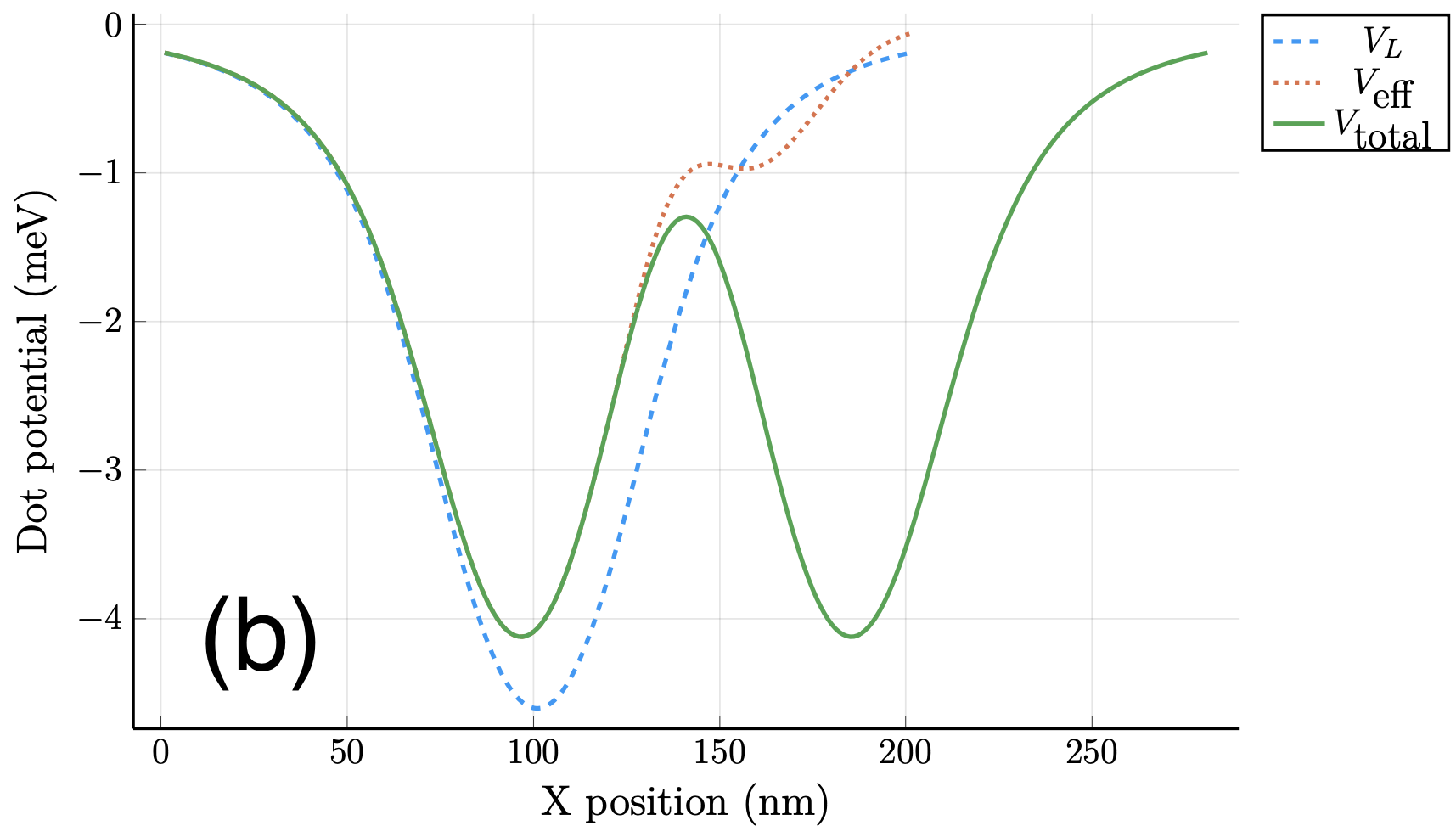}
      \label{fig:sfig2}
    \end{subfigure}
    \caption{A one-dimensional slice along the $x$ direction illustrating the effective single dot potential $V_\textrm{eff}$ that is equal to the combined potential $V_\text{total}$ times a filter function that tunes the potential to zero so that it remains centered on the first dot, as described in Eq.~\ref{eq:eff_pot}. \textbf{(a)} Plunger gate voltages $v_L, v_R = 0.2$ V, barrier gate voltage $v_X=0$ V. \textbf{(b)} Plunger gate voltages $v_L, v_R = 0.2$ V, barrier gate voltage $v_X=0.1$ V.} 
    \label{fig:eff_potential}
\end{figure*}

However, the single dot potential in Eq.~\ref{eq:dotPotential} is incomplete for dot-specific FCI calculations because these single dot potentials are unaffected by changes to the barrier gate.
In reality, we expect a lowered barrier gate to significantly increase wavefunction overlap between the two dots and that a raised gate would decrease that same overlap.  
We ensure that each single dot potential is both localized and accurately reflects the full potential---from both plunger gates and the barrier gate---at that site by taking the full potential and rapidly tuning it to zero as it approaches the second dot. 
\begin{equation}
  \label{eq:eff_pot}
 V_\textrm{eff}(x,y) = V_\text{total}(x,y) \exp\left[\frac{-\max(0,(x - d/4))^2}{(d/2)^2}\right]
\end{equation}
where 
\begin{equation}
  \label{eq:tot_pot}
  V_\text{total}(x,y) = V_L(x,y) - V_X(x,y) + V_R(x,y).
\end{equation}
This approach is illustrated in Fig.~\ref{fig:eff_potential} and we refer to these adjusted dots as effective single dot potentials.  
Separating the intradot and interdot interactions allows this approach to be applied to longer spin chains in principle, though we restrict our focus to double quantum dots in this work.

% Expand on this process. Include an equation

\begin{figure*}
  \begin{subfigure}{.5\textwidth}
    \centering
    \includegraphics[width=0.9\linewidth]{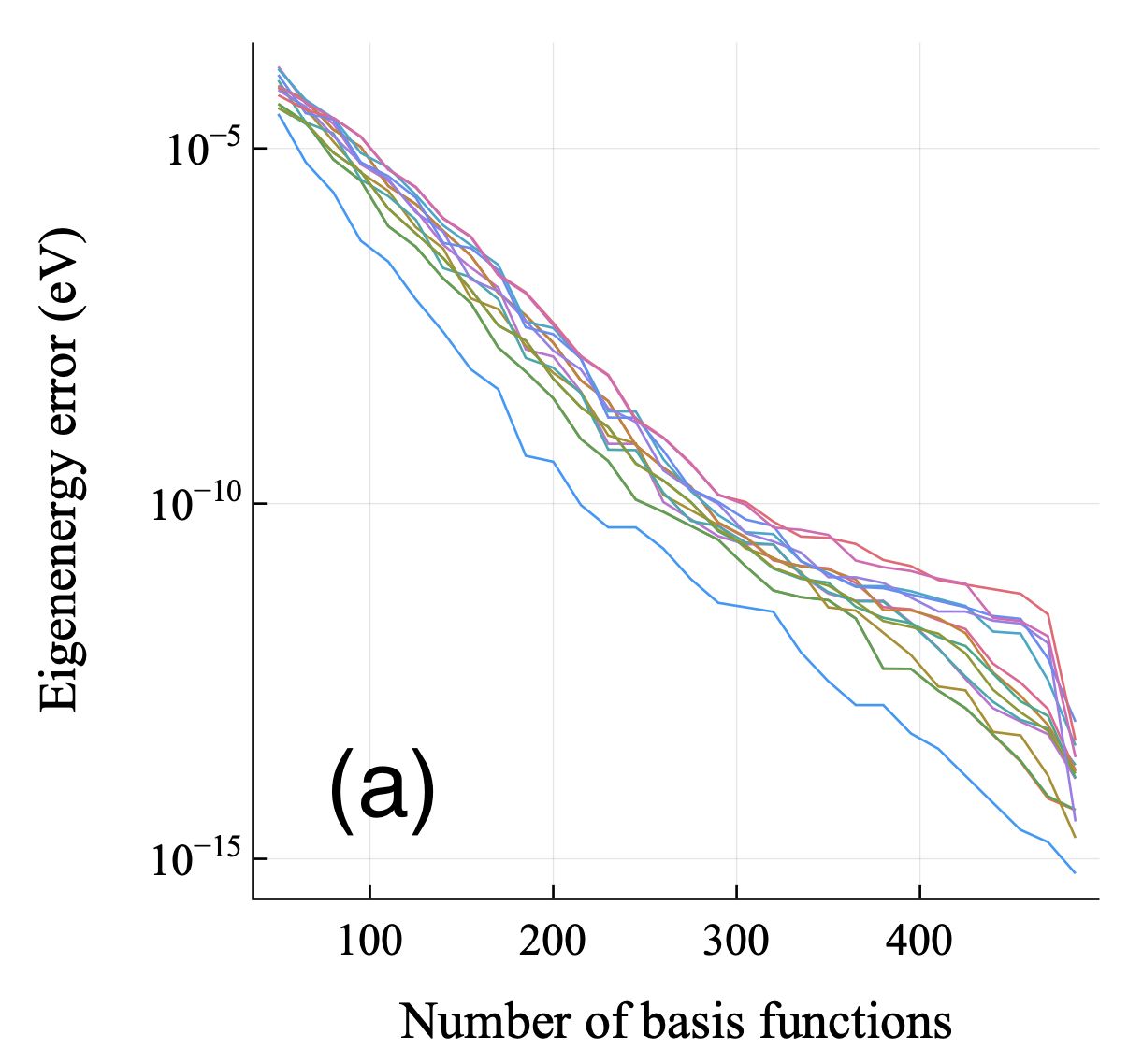}
  \end{subfigure}%
  \!\!\!\!\!\!\!\!\!\!\!\!\!\!\!\!\!\!\!\!\!\!\!
  \begin{subfigure}{.5\textwidth}
    \centering
    \includegraphics[width=0.8\linewidth]{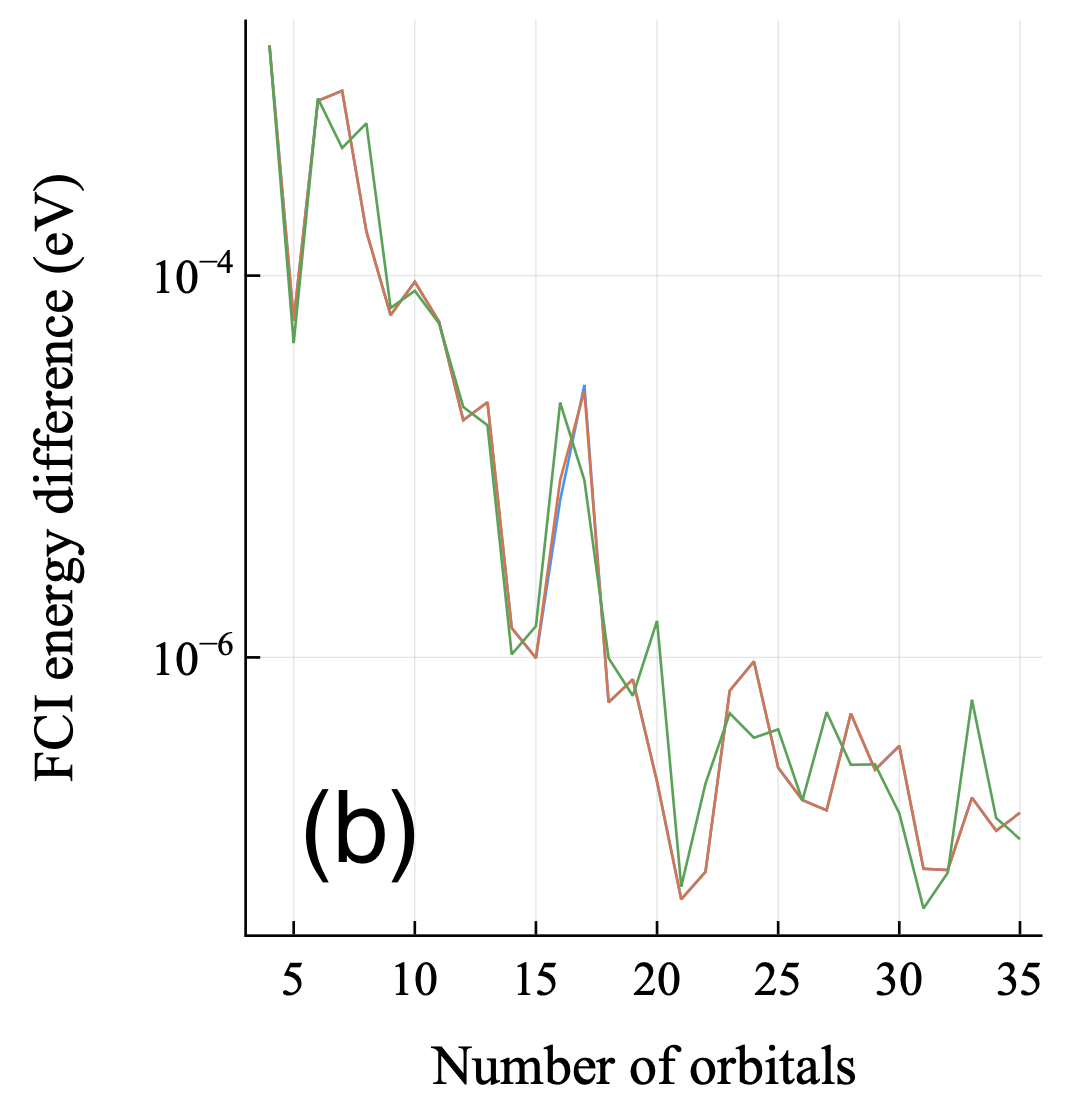}
  \end{subfigure}
  \caption{\textbf{(a)} The error in the eigenenergies of the first 15 orbitals. To compute the error, the energies were compared to their corresponding eigenenergy with 500 basis states. \textbf{(b)} The error in the FCI energies of the first 3 FCI states. To compute the error, the energies were compared to their corresponding FCI energies with 36 basis orbitals. }
  \label{fig:convergence}
\end{figure*}

Configuration interaction (CI) methods require expressing the many-body Hamiltonian in the basis of Slater determinants.
Each Slater determinant represents a different possible electron configuration, and the Coulomb term in Eq.~\ref{eq:fci} couples different configurations, hence the term \textit{configuration interaction}.
Given $M$ orbitals and $N$ electrons, there are $K = \binom{2M}{N}$ possible determinants.
The process is described as \textit{full} configuration interaction when all $K$ determinants are employed.
Given the fact that the number of determinants $K$ grows very quickly, we carefully choose our orbitals to keep the computational cost within reason and avoid memory constraints.
Rather than use a predetermined basis of orbitals for all calculations, we use the eigenstates of the single-particle Hamiltonian in Eq.~\ref{eq:fci}.
This approach has already been used effectively for spin qubit modeling of single electron dots by Anderson \textit{et al.}~\cite{anderson_high-precision_2022}.
This allows us to minimize the number of orbitals $M$ which must be kept.
Therefore, for each choice of gate voltages and pitch, we solve for the single-particle eigenstates of each dot using exact diagonalization in a truncated Fourier basis consisting of $N_F$ basis functions.
The eigenstates are approximated well after about $N_F = 500$ basis functions, and the eigenenergy convergence is shown in Fig.~\ref{fig:convergence}a.

Including enough orbitals $M$ is a critical part of a successful calculation, and we demonstrate the FCI ground state energy convergence in Fig.~\ref{fig:convergence}b.
Increasing $M$ until the output is consistent and predictable is mandatory (and is, in fact, the definition of self-consistent FCI), as it is very difficult to know \textit{a priori} how many orbitals will be necessary for sufficiently converged Hubbard parameters.
Ground state convergence is not by itself sufficient for converged Hubbard terms, however.
Separating out parts of a system's wavefunction and projecting to the Hubbard space requires higher-order terms arising from the excited states to be included. 
Early truncation can lead to erratic, numerically unstable values of the Hubbard parameters.
We have found keeping the first $M=30$ Slater determinants to be good practice for our choice of parameters, but this has to be ensured in each calculation since, depending on the details, more or fewer orbitals may be necessary for the computation.

Next, we need to extract the effective single-particle states from the FCI wavefunction.
For a single Slater determinant, this is simple.
However, such an approach would ignore correlational effects within the dot.
Separable approaches to the many-body system will never capture dynamical correlations, but important contributions from exchange and static correlation can still be incorporated.
Fortunately, when $N=3$, it becomes possible to isolate the valence electron well in such a way as to ensure that the two core electrons resemble a singlet state as much as possible.
This process consists of taking the one-particle reduced density matrix (1RDM)~\cite{szabo_modern_1982}, which is given as
\begin{equation}
  \gamma_{ij} = \langle \Psi | \hat{a}_i^\dagger \hat{a}_j | \Psi \rangle,
\end{equation}
where $| \Psi \rangle$ is the ground state FCI wavefunction.

The eigenvectors of this density matrix represent the natural orbitals of the system, and the eigenvalues are their occupation numbers.
Because there are three electrons in the dot, one of them can be isolated by its spin, which without loss of generality, we classify as spin up.
The natural orbitals that are spin up, and their occupation numbers represent one of the core electrons, and we construct a density matrix to describe its state.
We then can use the remaining spin-down orbitals to construct another state that resembles the core electron as best as possible. 
The remaining state is that of the valence electron.
This is similar to the occupation number representation of a Slater determinant since it is not important which electron interacts with which.
Only a correct understanding of the states and their occupation is required.
Consequently, the many-body state is separated into three single-particle operators.
Each single-particle density operator's spin is also well-defined, which maps well to the Hubbard representation.

\begin{figure}
  \includegraphics*[width=0.9\linewidth]{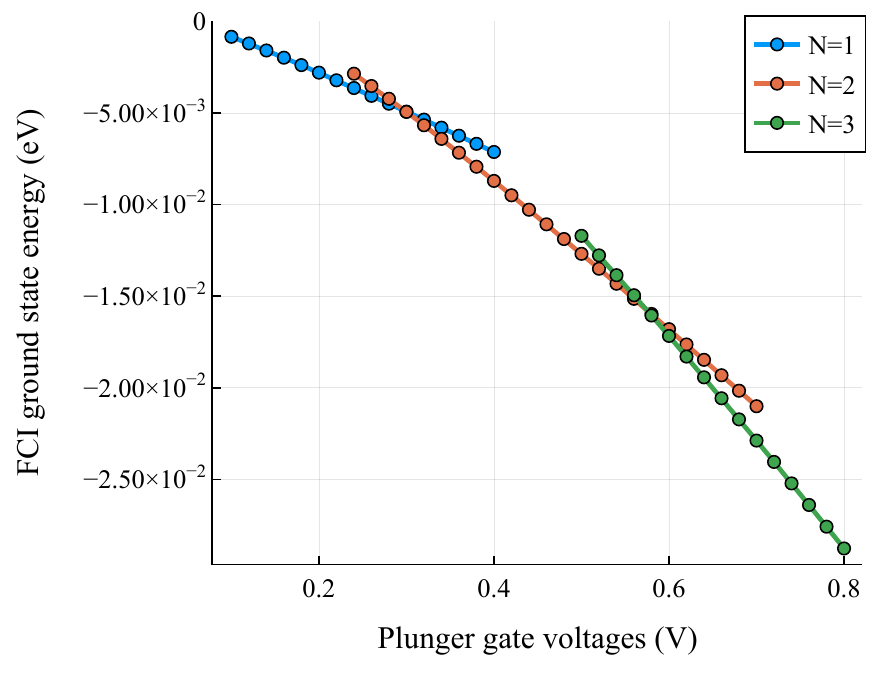}
  \caption{Multi-electron FCI ground state energies for the left dot as a function of plunger gate.
Both the left and right plunger gates are kept at the same voltage throughout. $d=80$ nm, $v_X = 0.1$ V}, and $B=100$ mT. 
  \label{fig:groundstates}
\end{figure}

These single-particle density operators then need to be orthogonalized with respect to the same operators on a neighboring dot since they will have a nonzero overlap for any finite pitch.
This is usually done by performing a L{\"o}wdin orthogonalization process, which is symmetric with respect to its arguments.
Under the L{\"o}wdin procedure, the vectors are rotated so that no single vector is rotated more than the others.
This way, all new vectors will resemble their original starting points as much as possible. 
This is the standard orthogonalization routine in quantum chemistry~\cite{szabo_modern_1982}.
L{\"o}wdin orthogonalization differs from the more familiar Gram-Schmidt process, which is strongly asymmetric in its treatment of the initial vectors.
However, a direct application of the L{\"o}wdin procedure is obviously inappropriate, since density matrices do not form a vector space. 
Therefore, we again work with the natural orbitals of each density matrix.

To orthogonalize our set of single particle density matrices, we use the density operator Uhlmann fidelity $F(\rho, \sigma)$ as the distance metric over which we define orthogonality.
\begin{equation}
 F(\rho,\sigma) = \text{tr}\left(\sqrt{\sqrt{\sigma} \rho \sqrt{\sigma}}\right)
\end{equation}
Fidelity between two density operators $F(\rho,\sigma) = 0$ if and only if $\rho$ and $\sigma$ have orthogonal support \cite{nielsen_quantum_2010}.
L{\"o}wdin orthogonalization on the eigenvectors of the single-particle density operators produces density matrices of single-particle states on neighboring dots with zero fidelity.

This orthogonalization is done as follows: We perform a change of basis from the $2M$ basis orbitals ($M$ from each dot), which are not mutually orthogonal, to orthogonalized versions of these basis states.
This is a passive transformation to place density matrices from neighboring dots on equal footing and to work in an orthonormal basis. 
We then find the natural orbitals that contribute most to each density matrix and orthogonalize them with respect to each other using the L{\"o}wdin procedure.
Orbitals are iteratively included until overlaps of states from neighboring dots reach a set threshold $\varepsilon = 1\times10^{-10}$. 
This is done because as more (irrelevant) orbitals are included in the orthogonalization procedure, the resulting states resemble their original versions less and less.
This lets us directly extract the relevant Hubbard parameters for obtaining the charge stability diagram. 
We need only to modify the initial equations to handle density matrices rather than wavefunctions.
\begin{equation}
 F_{ij} = \text{tr}\sqrt{\sqrt{\rho_i} \hat{H} \rho_j \hat{H} \sqrt{\rho_i}}
\end{equation}
\begin{equation}
 G_{ijkl} = \text{tr}\sqrt{\sqrt{\rho_i \otimes \rho_j} \hat{H} \rho_k \otimes \rho_l \hat{H} \sqrt{\rho_i \otimes \rho_j}}
\end{equation}
where $\rho_i$ corresponds to the state $c^\dagger_i |0\rangle$ in the Hubbard model.
With these quantities, we can efficiently simulate the charge stability diagram for a given gate architecture.

\section{Results}

\begin{figure*}
    \includegraphics*[scale=0.55]{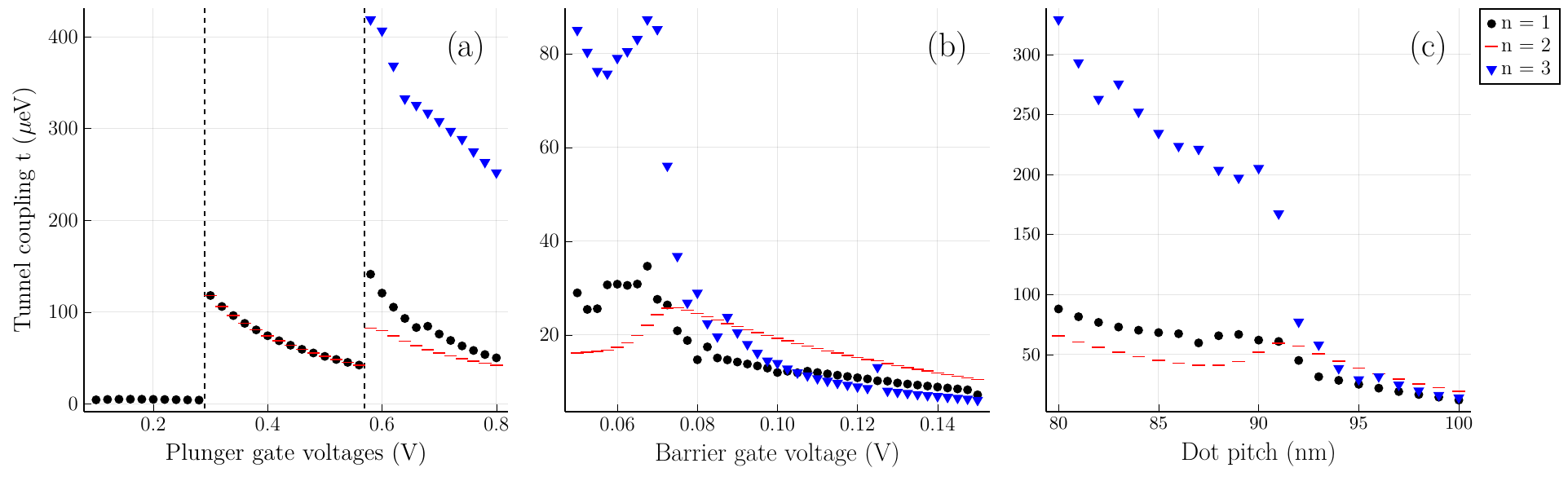}
    \caption{\textbf{(a)} Tunnel coupling $t$ as a function of the plunger gate voltage of both dots. Tunnel coupling is significantly enhanced for electrons in $p$-orbitals of each dot. $v_X$ = 0.1 V, $d$ = 80 nm, $B$ = 100 mT, and $M = 30$. \textbf{(b)} Tunnel coupling $t$ as a function of the barrier gate voltage. $v_i$ = 0.65 V, $d$ = 100 nm, $B$ = 100 mT, and $M = 30$. \textbf{(c)} Tunnel coupling $t$ as a function of the dot pitch. $v_X$ = 0.1 V, $v_i$ = 0.65 V, $B$ = 100 mT, and $M = 30$.}
    \label{fig:tunnelcouplings}
\end{figure*}

We take a one-dimensional slice in the charge stability diagram along the $v_1 = v_2$ diagonal and calculate the FCI ground state of each dot for $N=1,2,$ and $3$ to find the correct electron number. 
These results are shown in Fig.~\ref{fig:groundstates}.
The charge transitions of $V = 0.30$ and $V=0.58$ mark the boundaries of the charge regimes which can be used for further calculations. These transition points will depend on gate layout and barrier gate strength but are relatively cheap to determine for $N \le 3$ electrons since the ground state energy converges quickly in $M$ compared to Hubbard values.

We first calculate the tunnel coupling as a function of plunger gate voltage, barrier gate voltage, and dot pitch.
Fig.~\ref{fig:tunnelcouplings} displays our results.
We find that tunnel coupling varies significantly over different orbital states and that $t_3$, which corresponds to accessing the first $p$-orbital of each dot, sees a ~50x increase in tunnel coupling strength relative to the $N=1$ case for modest values of pitch and barrier gate.
This is sensible since the electrons are less localized to their original dot as they access higher orbital states, and we expect to see the overlap increase exponentially in such a regime. 
By ``$p$-orbital'' we mean the first excited orbital state of the quantum dot, since this state has odd parity and is analogous to the $p$-orbital of conventional atoms.
It is worth noticing that the value of $t$ fluctuates significantly within each charge state, with a monotonically decreasing behavior as the plunger gates are increased.
This is physically logical, as we would expect the states to become more localized. 
However, for the purposes of modeling charge stability diagrams, and specifically the transitions between charge states, this effect is minor compared to the order-of-magnitude shifts that occur at the charge transitions.

The effects of the barrier gate are apparent, with increased values of $v_X$ immediately diminishing the result of $t_3 \gg t_1$. 
This behavior is only observable because of our effective single dot potentials, which take into account the barrier gate strength in calculating the center of the single dot well. 
As we increase the barrier voltage, it appears that tunnel couplings for the $n=3$ electron drop below that of the core electrons. 
This is likely only a numerical artifact, as the tunnel couplings in that regime are small enough that our separation and orthogonalization procedure may fail to capture the correct (but very small) value. 
Numerical errors in such exponentially small tunnel coupling values are not a problem for obtaining accurate charge stability diagrams.

Analyzing the tunnel couplings calculated as the dot pitch is varied (Fig.~\ref{fig:tunnelcouplings}) also illustrates that this regime of very large $t_\textrm{valence}$ is quickly suppressed for larger pitches or barriers. Here it appears that $d = 90$ nm is a key value (for our parameters), after which the tunnel coupling starts to drop precipitously. Similar calculations could be useful for estimating the spatial distribution of the valence electrons in individual devices.

\begin{figure*}
  \begin{subfigure}{.5\textwidth}
    \centering
    \includegraphics*[width=0.9\textwidth]{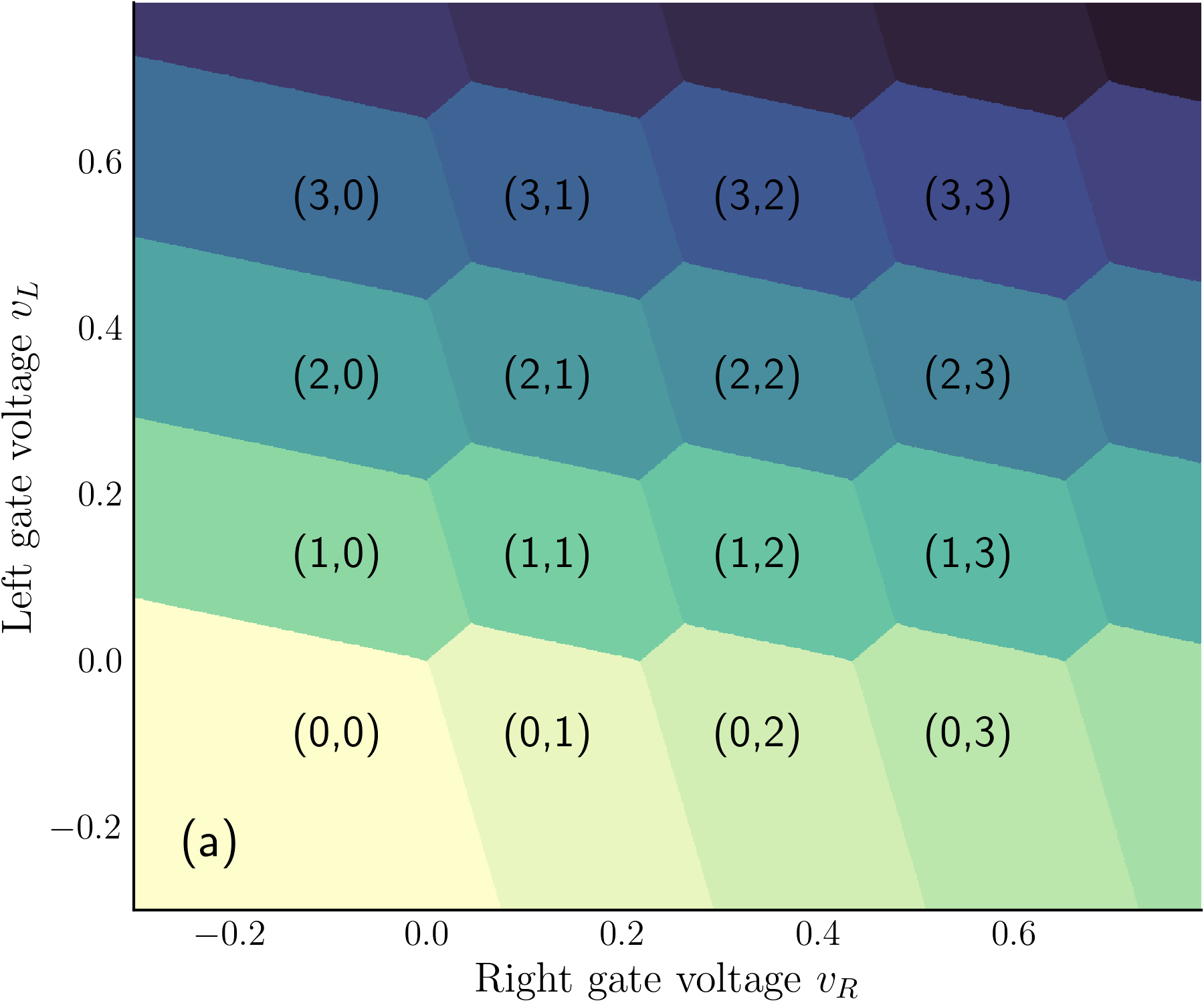}
  \end{subfigure}%
  \!\!\!
  \begin{subfigure}{.5\linewidth}
    \centering
    \includegraphics*[width=0.9\textwidth]{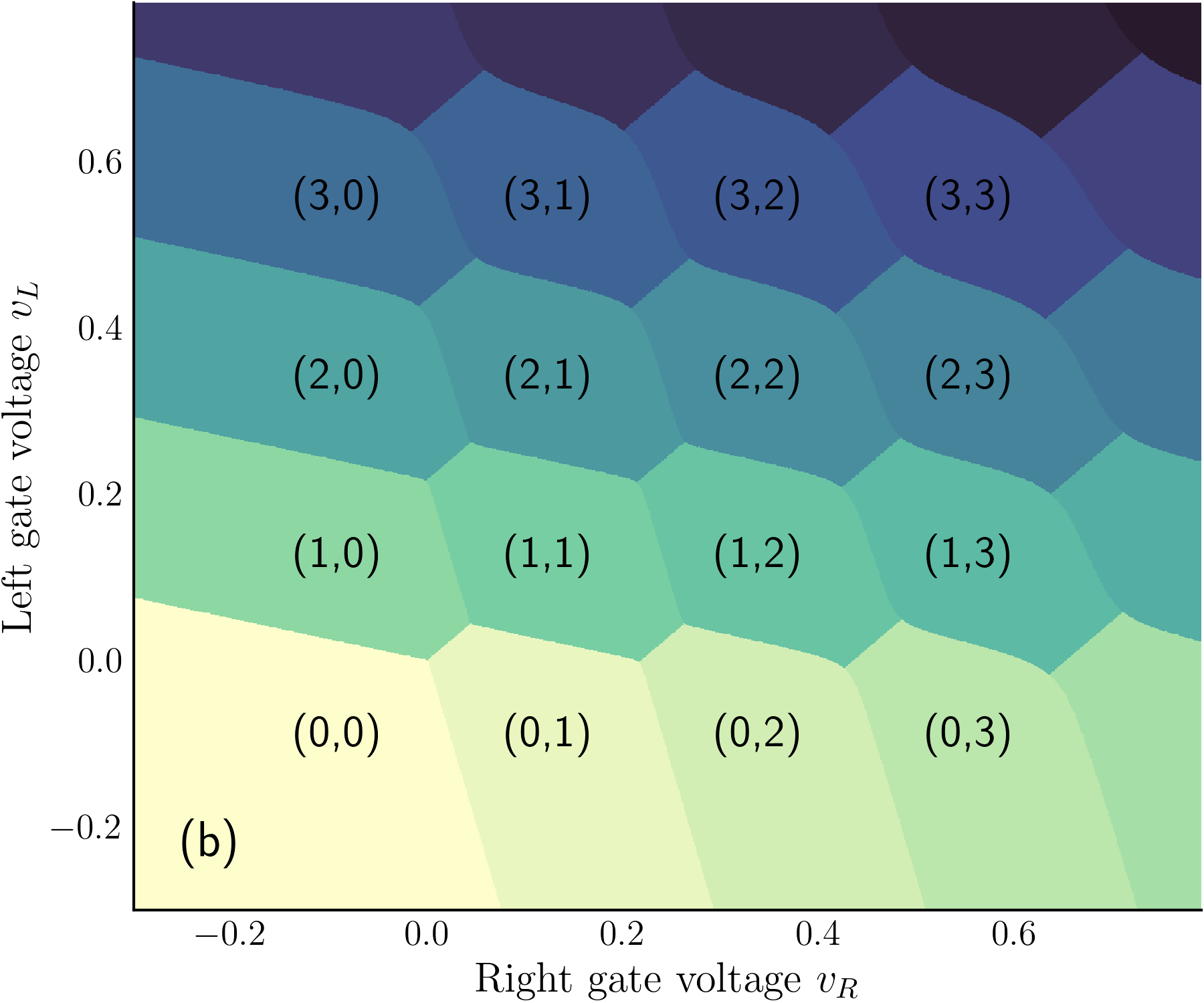}
  \end{subfigure}
  \caption{A pair of charge stability diagrams for $v_X = 0.1$ V, $d=80$ nm. The diagram on the left demonstrates the standard approach of using a level-independent value of $t_i = 40$ $\mu$eV. For both plots, $U=4.0$ meV and $V = 1.0$ meV. The plot on the right has tunnel couplings $t_i = [10, 80, 420]$ $\mu$eV. Larger tunnel coupling leads to greater smoothing in the higher occupancy Coulomb cells.}
  \label{fig:levdependence}
\end{figure*}
\begin{figure*}
  \begin{subfigure}{.5\textwidth}
    \centering
    \includegraphics*[width=0.9\textwidth]{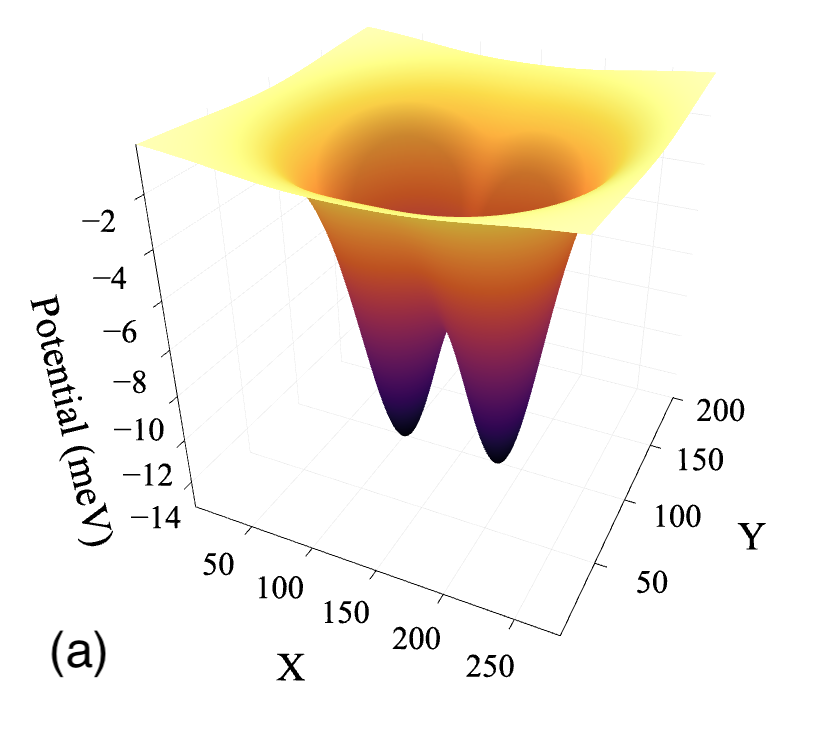}
  \end{subfigure}%
  \!\!\!\!\!\!\!\!\!\!
  \begin{subfigure}{.5\textwidth}
    \centering
    \includegraphics*[width=0.9\textwidth]{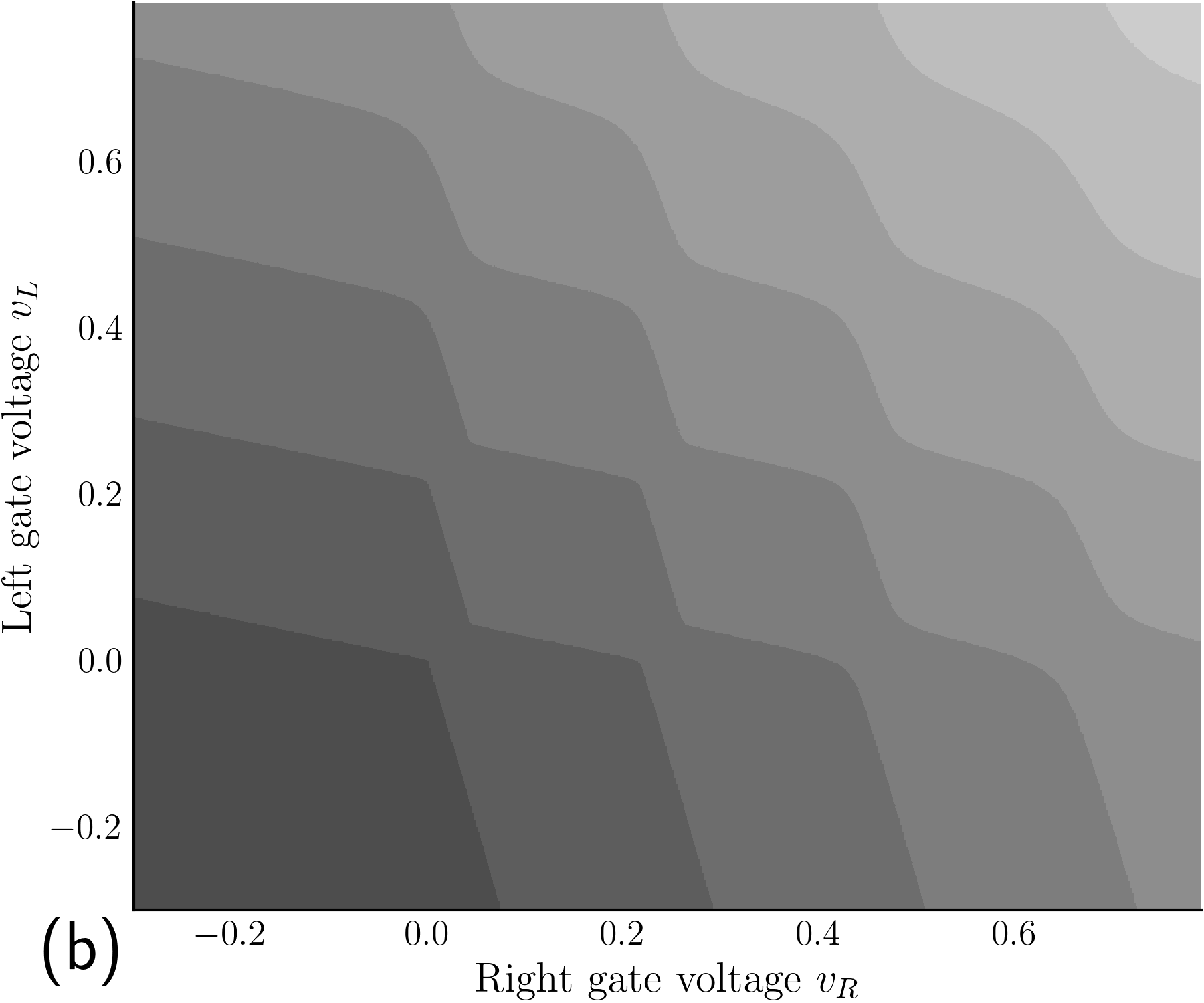}
  \end{subfigure}
  \caption{ \textbf{(a)} A three-dimensional view of the double dot potential used for the generation of Fig.~\ref{fig:levdependence}, with dot pitch $d = 80$ nm and barrier gate $v_X = 0.1$ V. For this image, the plunger gates are set at $v_i = 0.60$ V. \textbf{(b)} The same system visualized as a stability diagram of total charge.
 }
  \label{fig:levdependence_total}
\end{figure*}

The effect of these occupancy-dependent tunnel couplings is seen when generating the charge stability diagram.
In Fig.~\ref{fig:levdependence}a, we calculate a charge stability diagram using only Hubbard parameters calculated from a double dot system with plunger gates at $0.2$ V. 
The Hubbard parameters are applied over the whole stability diagram.  
Switching on level-dependent tunneling leads to the picture in Fig.~\ref{fig:levdependence}b, as the increased tunnel coupling leads to greater corner smoothing for higher charge states.
Because the major changes in $t$ occur at charge transitions, we model $t$ in the Hubbard model as being piecewise constant over each Coulomb diamond. This already represents a significant improvement over the widespread practice of assuming that the tunnel coupling is constant over the entire stability diagram. If a transition involves tunneling from one dot with $n$ electrons into another with $m$ electrons, the tunnel coupling corresponding to that interaction is $t_j$, where $j = \max(n,m+1)$. 
Charge stability diagrams are not the only experimental measurements that reflect occupancy-dependent tunneling. We would expect direct measurements of the interdot exchange interaction energy to reflect this occupancy-dependent tunneling as well.
This occupancy-dependent tunneling is a new qualitative feature of our theory compared with the existing charge stability theories in the literature.

Because of the dramatically higher levels of tunnel coupling, points on the charge stability diagram that are one charge state in Fig.~\ref{fig:levdependence}a could be another one entirely in Fig.~\ref{fig:levdependence}b. 
Tuning quantum circuits for quantum-dot-based qubits depends critically on the exact locations of charge boundaries, and therefore differences in Fig.~\ref{fig:levdependence} that appear small to the eye can have far-reaching effects when it comes to gate fidelities.

\begin{figure}
  \centering
    \hspace{-0.2\linewidth}
    \includegraphics*[width=0.75\linewidth]{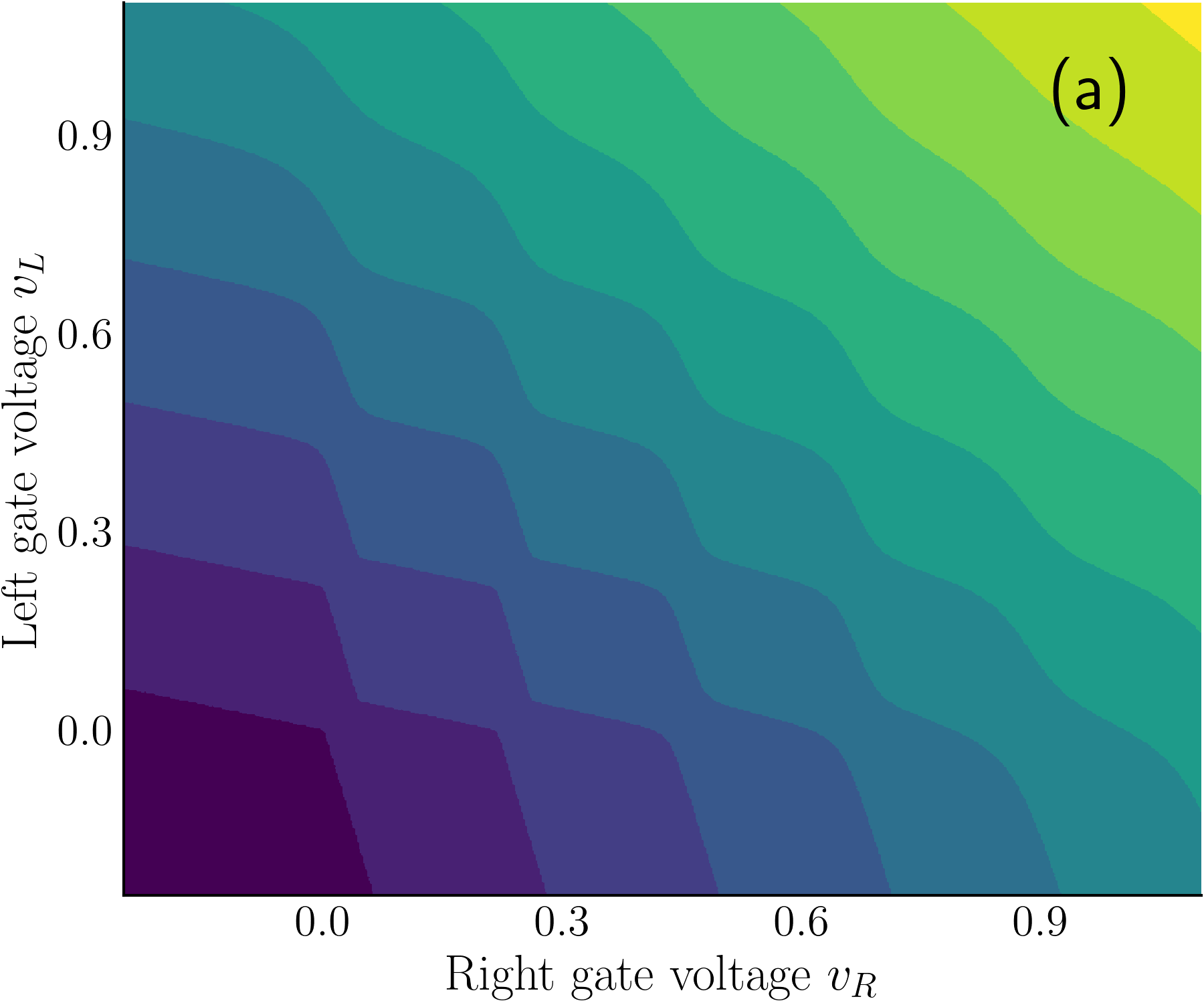} \\
    
    \includegraphics*[width=0.98\linewidth]{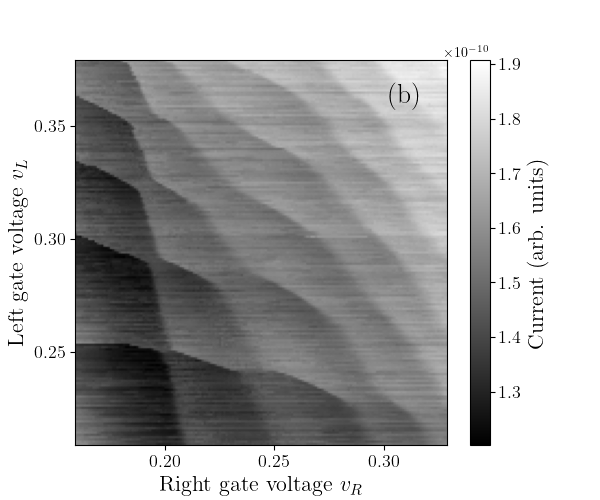}
  \caption{\textbf{(a)} An extrapolated charge stability diagram, where we seek to demonstrate the effects of the trend of increasing tunnel couplings as occupancy is increased. We generate this plot by simulating the Hubbard model with up to 6 electrons in each dot. $U=4.0$ meV, $V=1.0$ meV, and $t_i = [0.01, 0.08, 0.42, 0.8, 1.6, 2.0]$ meV. \textbf{(b)} An example of an experimental charge stability diagram exhibiting occupancy-dependent tunneling, with high occupancy bands becoming increasingly featureless. This diagram is contained in the QFlow dataset~\cite{zwolak_qflow_2018}.}
  \label{fig:n6}
\end{figure}

Another feature in the level-dependent coupling diagram is that for $N > 3$, the individual Coulomb diamonds become less pronounced, and the diagonal slices corresponding to the double dot total charge become more straight and featureless (see Fig.~\ref{fig:n6}). 
This feature is often seen in the experimental data~\cite{zajac_resonantly_2018,ziegler_tuning_2023,zwolak_autotuning_2020}.
Experimental data usually only shows changes in the total charge of the system, though occasionally charge transitions at the dot level are also visible for a small dot occupation. Fig.~\ref{fig:levdependence_total}b reflects how level-dependent tunneling may appear in total charge transition diagrams.

\begin{figure*}
  \includegraphics*[scale=0.58]{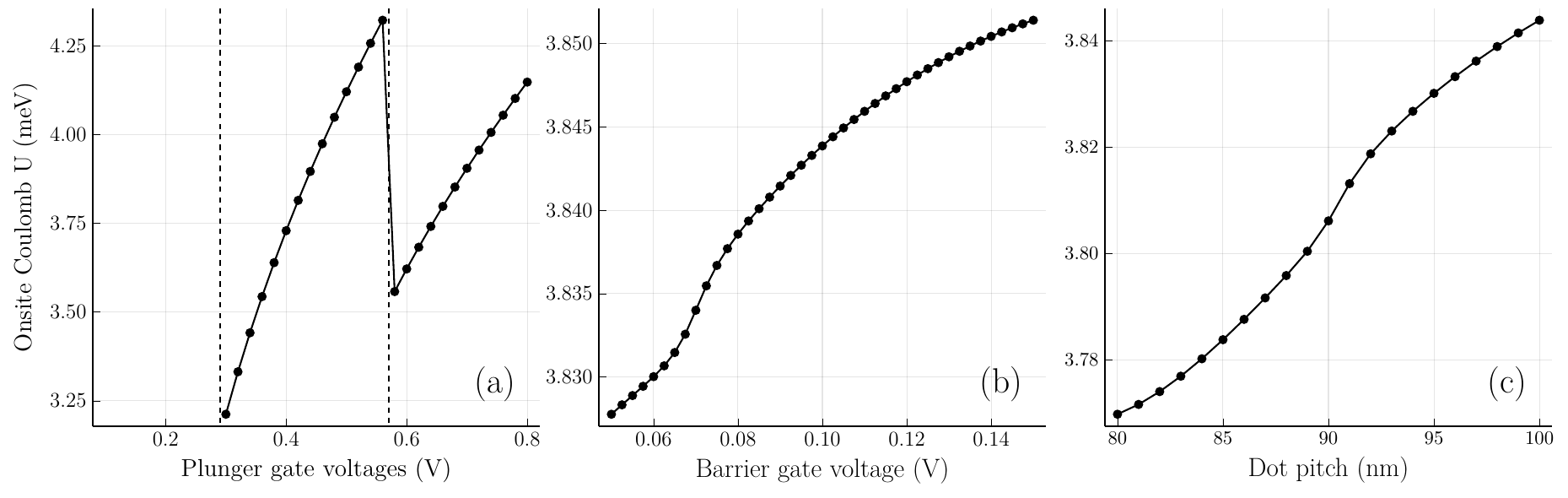}
  \caption{\textbf{(a)} Onsite Coulomb term $U$ as a function of the plunger gate voltage of both dots. $v_X$ = 0.1 V, $d$ = 80 nm, $B$ = 100 mT, and $M = 30$. There is no calculation of $U$ for $V < 0.30$ V  because the ground state has only one electron and therefore no Coulomb energy. \textbf{(b)} Onsite Coulomb term $U$ as a function of the barrier gate voltage. $v_i$ = 0.65 V, $d$ = 100 nm, $B$ = 100 mT, and $M = 30$. \textbf{(c)} Onsite Coulomb term $U$ as a function of the dot pitch. $v_X$ = 0.1 V, $v_i$ = 0.65 V, $B$ = 100 mT, and $M = 30$. All Coulomb terms $U$ plotted here are the average value of $U$ within a single dot. }
  \label{fig:us}
\end{figure*}
\begin{figure*}
  \includegraphics*[scale=0.58]{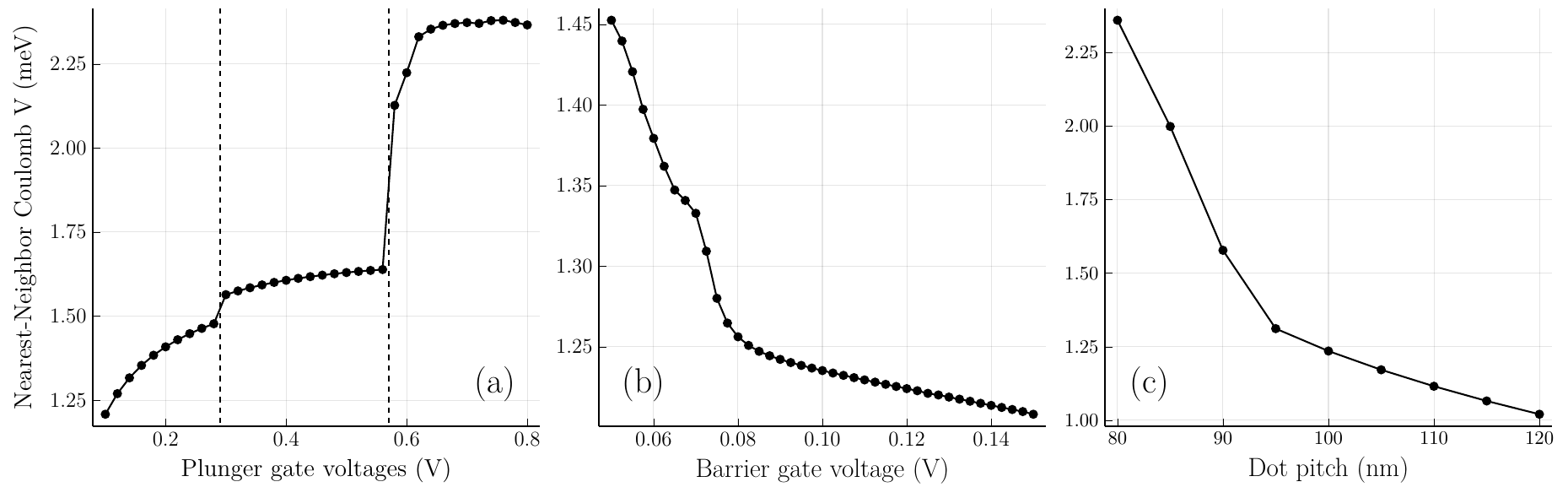}
  \caption{\textbf{(a)} Coulomb coupling $V$ as a function of the plunger gate voltage of both dots. $v_X$ = 0.1 V, $d$ = 80 nm, $B$ = 100 mT, and $M = 30$. \textbf{(b)} Coulomb coupling $V$ as a function of the barrier gate voltage. $v_i$ = 0.65 V, $d$ = 100 nm, $B$ = 100 mT, and $M = 30$. \textbf{(c)} Coulomb coupling $V$ as a function of the dot pitch. $v_X$ = 0.1 V, $v_i$ = 0.65 V, $B$ = 100 mT, and $M = 30$. All Coulomb couplings plotted here are calculated between the two dots' valence electrons.}
  \label{fig:vs}
\end{figure*}

We performed a similar analysis for the onsite Coulomb term $U$ in Fig.~\ref{fig:us} and the Coulomb coupling term $V$ in Fig.~\ref{fig:vs}.
The onsite Coulomb term $U$ is unique in that it does not couple states of neighboring dots like $t$ or $V$, and depends entirely on the properties of a single dot.
For that reason, it is calculated differently. Rather than using atomic orbitals to calculate the overlap of the charge distributions, we simply calculate the expectation value of the Coulomb Hamiltonian for our FCI single-dot ground state.
Fig.~\ref{fig:us} shows that the pitch and barrier gate have very small effects on $U$. However, as before with the tunnel coupling, we see significant fluctuations within a given charge state.
It is reasonable and even expected that the Coulomb energy increases as the multielectron wave function becomes more localized.
However, in keeping with our purpose of modeling charge transitions, $U$ is assumed to be constant over the charge stability diagram, since we do not see dramatic shifts in the average value of $U$ between Coulomb diamonds, unlike the tunnel coupling.

Our calculations of the Coulomb coupling term $V$ are shown in Fig.~\ref{fig:vs}. 
The Coulomb energy between valence electrons of neighboring dots appears to have large shifts at charge transitions and to remain relatively constant within each regime.
As the wells become deeper within a charge regime, we do not expect $V$ to change significantly, whereas we would expect larger changes to $V$ when the dot wavefunctions have larger overlaps, such as when $p$-orbitals of the dot are occupied, or the barrier gate or pitch is lowered.
This is exactly what we see in the results, with $p$-orbitals having significantly larger Coulomb couplings.
We also observe $V$ decreasing with barrier gate voltage and pitch in a manner generally similar to that of the tunnel coupling. 

It would appear that a piecewise constant approach to $V$ would be appropriate, similar to our approach with the tunnel coupling. However, even if such an approach is pursued, the relevant quantity is the average value of $V = \frac{1}{n_1 n_2}\sum_{i,j} V_{ij}$. We employ this approach, and our charge stability diagrams use a constant value of $V$ which is the average of the Coulomb couplings, since this average stays approximately the same across a single charge stability diagram.

\begin{figure*}
    \begin{subfigure}{.5\textwidth}
      \centering
      \includegraphics*[width=0.9\textwidth]{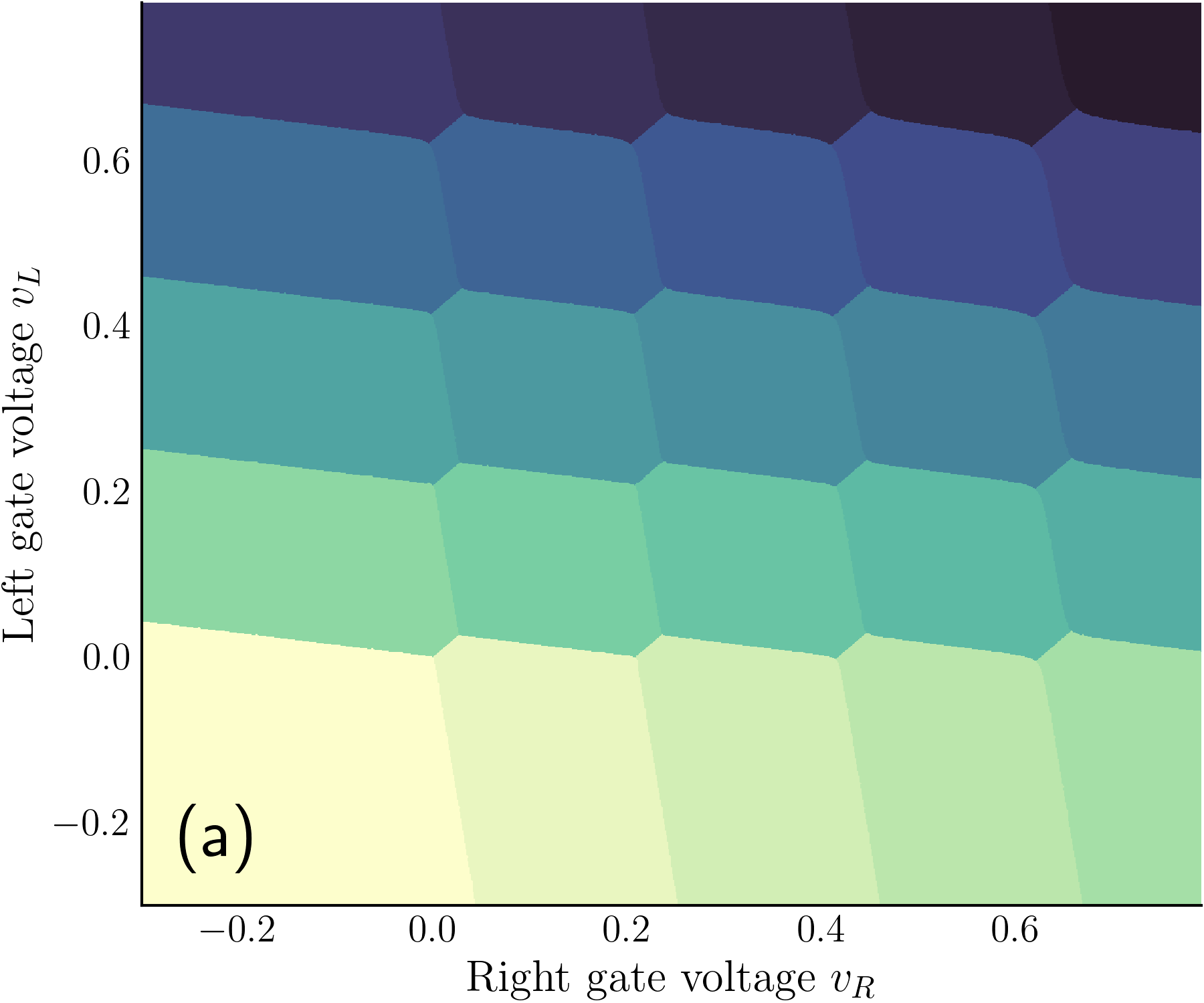}
    \end{subfigure}%
    \begin{subfigure}{.5\textwidth}
      \centering
      \includegraphics*[scale=0.48]{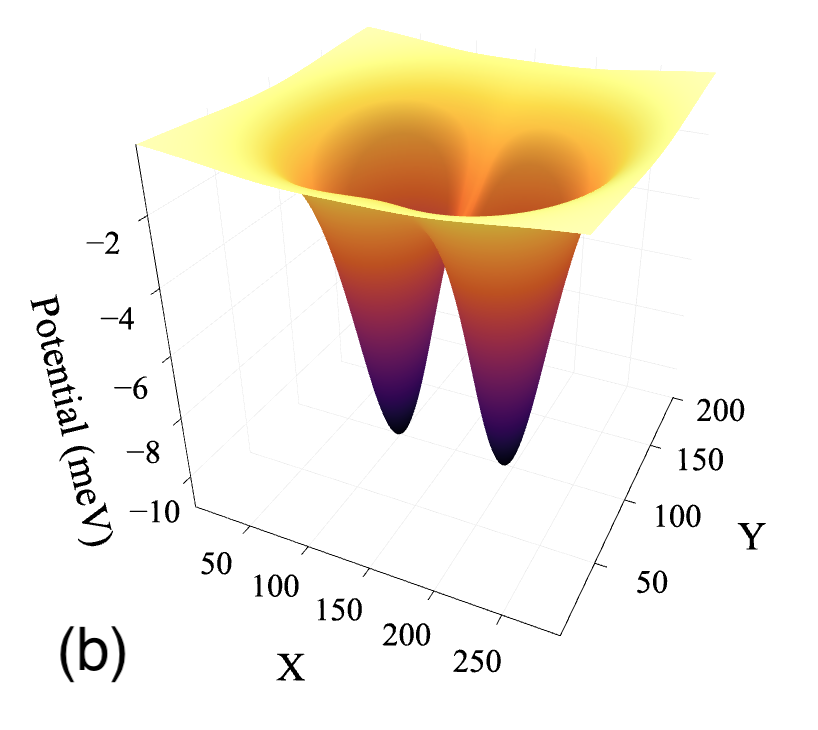}
      \vspace{10mm}
    \end{subfigure}
    \\
    \begin{subfigure}{.5\textwidth}
      \centering
      \includegraphics*[width=0.9\textwidth]{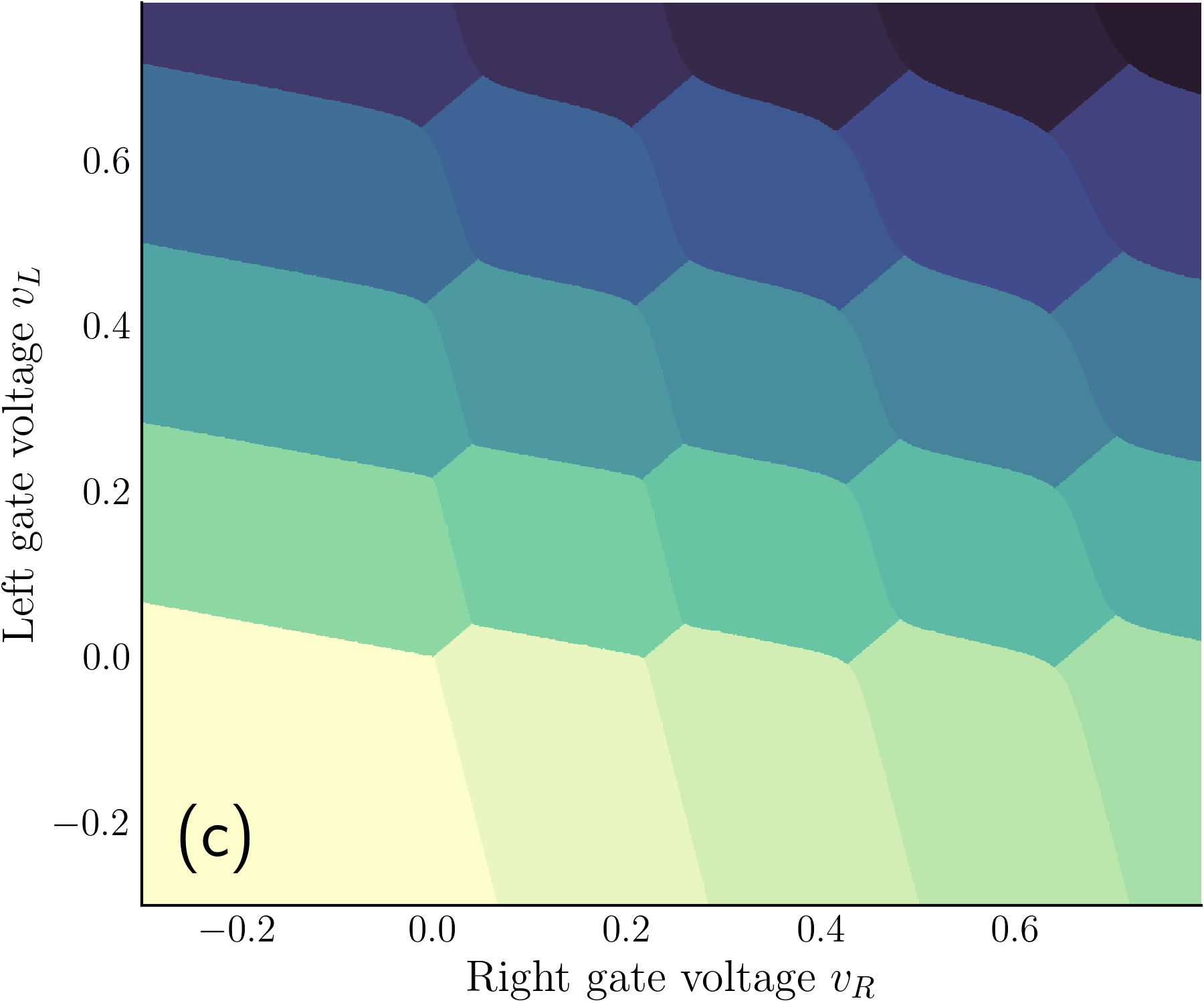}
    \end{subfigure}%
    \begin{subfigure}{.5\textwidth}
      \centering
      \includegraphics*[scale=0.48]{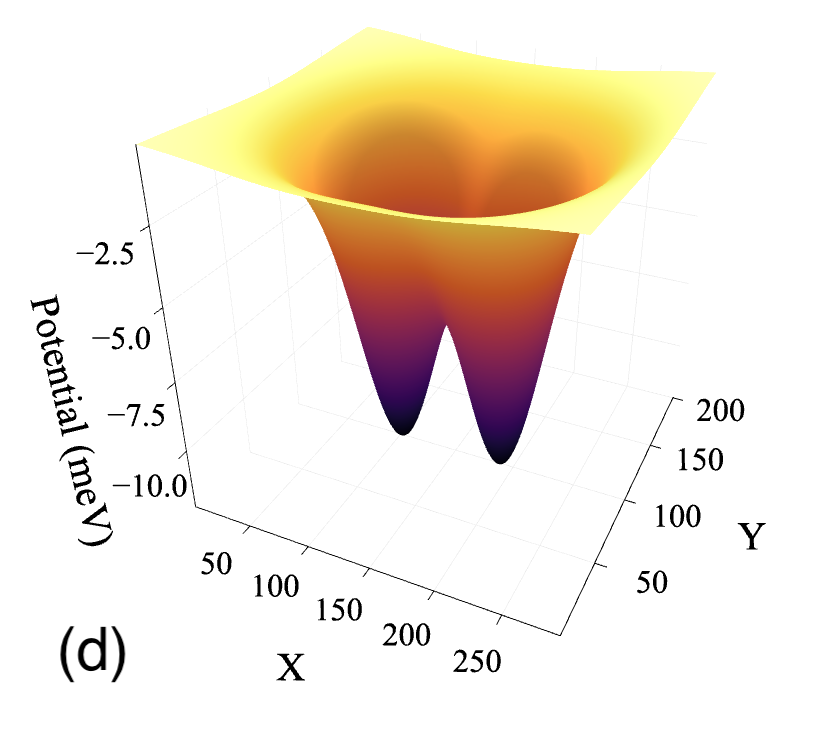}
      \vspace{10mm}
    \end{subfigure}%
    \caption{\textbf{(a)} A charge stability diagrams for $v_X = 0.15$ V, $d=90$ nm.  $U=4.2$ meV and $V = 0.6$ meV.  This system has tunnel couplings $t_i = [10, 35, 180]$ $\mu$eV \textbf{(b)} The corresponding background potential. \textbf{(c)} A charge stability diagrams for $v_X = 0.0$ V, $d=90$ nm.  $U=4.1$ meV and $V = 0.9$ meV.  This system has tunnel couplings $t_i = [10, 60, 380]$ $\mu$eV \textbf{(d)} The corresponding background potential.}
    \label{fig:barrierdependence}
\end{figure*}
\begin{figure}
    \centering
    \includegraphics*[scale=0.6]{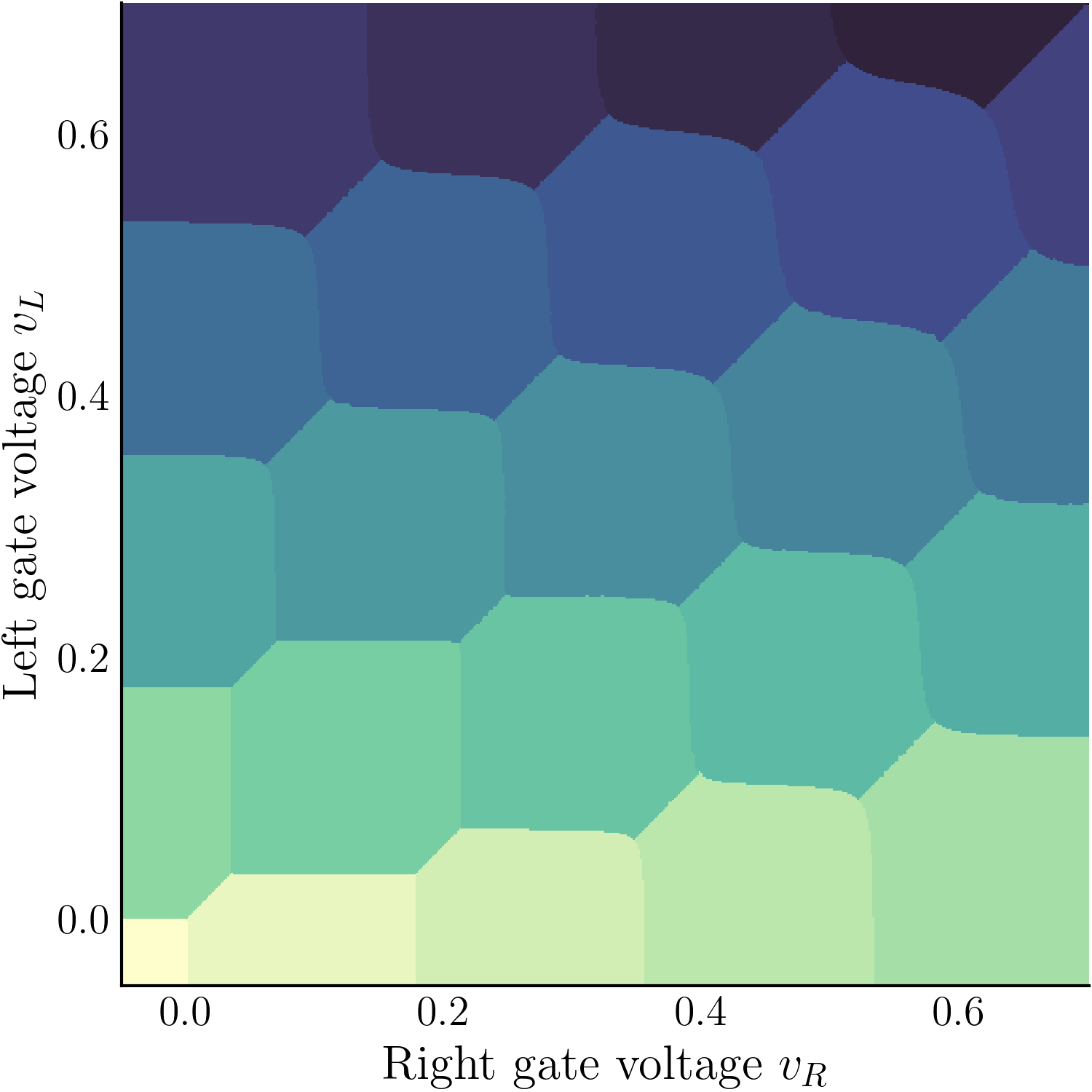}
    \caption{An example of a charge stability diagram in terms of virtual gate voltages. This process accounts for electrostatic crosstalk, and ``straightens out'' the stability diagram. This figure is generated with $U=4.1$ meV, $V=0.8$ meV, and $t_i = [10, 40, 400]$ $\mu$eV.}
    \label{fig:gatevirt}
\end{figure}

The tunnel coupling varies dramatically throughout the stability diagram, and still only has a small (but nonetheless important) effect on the final stability diagram. Therefore, it is to be expected that the changes to the Coulomb terms, both the intradot $U$ and the interdot $V$ repulsion, can be safely ignored \textit{within} a single stability diagram since the corrections in charge distribution that arise from accessing higher orbital states have only a minor effect on the Coulomb energies. Nonetheless, it is worth considering the effect that the barrier gate and dot pitch have on the entire charge stability diagram. We show these effects in Fig.~\ref{fig:barrierdependence}.

The fact that $V$ significantly increases as the barrier gate is lowered means that charge stability diagrams can look drastically different for the same gate geometry but with a slightly lower barrier gate voltage. Lowering barrier gate voltage also tends to decrease $U$ slightly, but the increase in $V$ is much more dramatic.  We again stress the importance of our effective single dot potentials for capturing these effects. Fig.~\ref{fig:barrierdependence} clearly demonstrates that given a situation with enhanced $t_3$ coupling, this behavior can be suppressed by raising the barrier gate, and thereby returning the charge stability map to one where $N=3$ Coulomb diamonds more closely resemble those corresponding to $N=1$.

Our charge stability plots are generated in real space, starting from the $(\mu_1, \mu_2)$ space and including neighboring dot occupation and electrostatic crosstalk through the interdot Coulomb effects with the standard formulas from the classical capacitance model.

\begin{eqnarray*}
    \mu_1 &= |e|\Bigl[\alpha_1 v_1 + (1-\alpha_1) v_2\Bigr] \\
    \mu_2 &= |e| \Bigl[(1-\alpha_2) v_1 + \alpha_2 v_2\Bigr] \\ 
    \alpha_1 &= (U_2 - V)U_1/(U_1 U_2 - V^2) \\
    \alpha_2 &= (U_1 - V)U_2/(U_1 U_2 - V^2)
\end{eqnarray*}

The alternative is to ignore these effects and plot the stability diagram using these theoretical quantities as the axes. 
These quantities are known as ``virtualized''.
Gate virtualization results in ``straightened out'' stability plots with horizontal or vertical charge boundaries. 
It is more meaningful for theorists to express the stability diagram results in physical units to better compare their work to experimental results since the virtualization procedure distorts the stability diagram in such a way that the effects of modeling choices can be exaggerated. 
It could also be beneficial for experimentalists to provide virtualized charge stability diagrams to better connect with theory. An example of a virtualized stability diagram is shown in Fig.~\ref{fig:gatevirt}.

\section{Conclusions}
In conclusion, we have presented a multi-band and multi-electron interacting model of gate-defined double quantum dots and simulated charge stability diagrams for these systems using realistic electrostatic potentials, full configuration interaction, and effective Hubbard model mapping.
Our approach is compatible with confining potentials of any shape and at any level of sophistication, including potentials obtained from Poisson-Schrodinger solvers as the starting point (replacing our model analytical confining potentials).
We separated the double dot system using an effective single dot approach that accounts for the barrier gate voltage.
We then utilized full configuration interaction to account for many-body interactions within each quantum dot.
Single particle density matrices were computed using natural orbitals of the 1RDM, and Hubbard couplings were calculated as the Uhlmann fidelities of the corresponding operators.
These parameters of the generalized Hubbard model enabled us to predict large regions of the charge stability diagram with improved accuracy compared to classical capacitance models or single-band Hubbard models. 

Our results highlighted the significant enhancement of tunnel couplings for valence electrons in dots with three electrons compared to single-electron dots. 
This difference was evident in the charge stability diagram for higher occupancy Coulomb diamonds. 
We extrapolate this trend to a large charge stability diagram in Fig.~\ref{fig:n6}.
As dot occupancies increase, so do the tunnel couplings, and total charge regions become increasingly featureless and eventually become almost parallel lines.
Unlike the tunnel coupling, the average values of the onsite Coulomb term $U$ and the interdot Coulomb coupling $V$ were not found to vary significantly within the charge stability diagram.
We also found that increasing barrier strength or dot pitch reduces the effect of increased tunnel coupling, and this trend allows researchers to make their Coulomb diamonds as periodic as possible so that they see similar behavior using $N=3$ dots as they do $N=1$ dots.

Our FCI approach is only applicable in regimes where it is appropriate to treat the dots as separate subsystems, where wavefunction overlap is small. 
Double dot systems with large detunings or negative barrier gates are not compatible with this technique.
Likewise, calculations of interdot exchange interaction rely on extremely precise pictures of the correlational energy of the entire double dot system, and cannot be performed in this formalism, even when tunnel couplings are small. 
Although calculating the FCI energy of the double dot system as a whole is an excellent approach for such a problem~\cite{anderson_high-precision_2022}, this would likely be computationally feasible only for $N<3$ electrons in each dot.

This work opens the door to many exciting future research directions, such as how background disorder in the form of charge noise or magnetic impurities impacts the Hubbard parameters.
Evaluations of the resulting charge stability diagrams could be done by calculating the ground state charge states of the double dot system explicitly using FCI or another quantum chemistry approach. Although FCI would yield an extremely accurate answer, it probably could only be used for the first few Coulomb diamonds because of computational cost.
One advantage of this approach is that because it separates the dots, it could feasibly be applied to spin chains longer than two to investigate how next-nearest and next-next-nearest neighbor tunnel couplings scale compared to nearest neighbor couplings.

Our work demonstrates the importance of considering multi-electron effects in double quantum dot systems and the potential for the generalized Hubbard model to capture these effects. 
By accurately predicting the charge boundaries of the stability diagram, our approach provides valuable insights for tuning multidot devices and achieving specific electron number states. 
We believe that this technique will allow for cost-effective simulation of very high-quality charge stability diagrams, which should be relevant for constructing gating sequences, prototyping device designs, and generating data for machine learning algorithms. 
Overall, our findings significantly contribute to the understanding of charge stability diagrams of quantum dot semiconductor spin qubits and enhance their potential for scalable quantum computing applications. 

\section{Acknowledgement}
This work was supported by the Laboratory for Physical Sciences.

\bibliography{references}

\begin{thebibliography}{38}%
\makeatletter
\providecommand \@ifxundefined [1]{%
 \@ifx{#1\undefined}
}%
\providecommand \@ifnum [1]{%
 \ifnum #1\expandafter \@firstoftwo
 \else \expandafter \@secondoftwo
 \fi
}%
\providecommand \@ifx [1]{%
 \ifx #1\expandafter \@firstoftwo
 \else \expandafter \@secondoftwo
 \fi
}%
\providecommand \natexlab [1]{#1}%
\providecommand \enquote  [1]{``#1''}%
\providecommand \bibnamefont  [1]{#1}%
\providecommand \bibfnamefont [1]{#1}%
\providecommand \citenamefont [1]{#1}%
\providecommand \href@noop [0]{\@secondoftwo}%
\providecommand \href [0]{\begingroup \@sanitize@url \@href}%
\providecommand \@href[1]{\@@startlink{#1}\@@href}%
\providecommand \@@href[1]{\endgroup#1\@@endlink}%
\providecommand \@sanitize@url [0]{\catcode `\\12\catcode `\$12\catcode
  `\&12\catcode `\#12\catcode `\^12\catcode `\_12\catcode `\%12\relax}%
\providecommand \@@startlink[1]{}%
\providecommand \@@endlink[0]{}%
\providecommand \url  [0]{\begingroup\@sanitize@url \@url }%
\providecommand \@url [1]{\endgroup\@href {#1}{\urlprefix }}%
\providecommand \urlprefix  [0]{URL }%
\providecommand \Eprint [0]{\href }%
\providecommand \doibase [0]{https://doi.org/}%
\providecommand \selectlanguage [0]{\@gobble}%
\providecommand \bibinfo  [0]{\@secondoftwo}%
\providecommand \bibfield  [0]{\@secondoftwo}%
\providecommand \translation [1]{[#1]}%
\providecommand \BibitemOpen [0]{}%
\providecommand \bibitemStop [0]{}%
\providecommand \bibitemNoStop [0]{.\EOS\space}%
\providecommand \EOS [0]{\spacefactor3000\relax}%
\providecommand \BibitemShut  [1]{\csname bibitem#1\endcsname}%
\let\auto@bib@innerbib\@empty
%</preamble>
\bibitem [{\citenamefont {Hanson}\ \emph {et~al.}(2007)\citenamefont {Hanson},
  \citenamefont {Kouwenhoven}, \citenamefont {Petta}, \citenamefont {Tarucha},\
  and\ \citenamefont {Vandersypen}}]{hanson_spins_2007}%
  \BibitemOpen
  \bibfield  {author} {\bibinfo {author} {\bibfnamefont {R.}~\bibnamefont
  {Hanson}}, \bibinfo {author} {\bibfnamefont {L.~P.}\ \bibnamefont
  {Kouwenhoven}}, \bibinfo {author} {\bibfnamefont {J.~R.}\ \bibnamefont
  {Petta}}, \bibinfo {author} {\bibfnamefont {S.}~\bibnamefont {Tarucha}},\
  and\ \bibinfo {author} {\bibfnamefont {L.~M.~K.}\ \bibnamefont
  {Vandersypen}},\ }\bibfield  {title} {\bibinfo {title} {Spins in few-electron
  quantum dots},\ }\href {https://doi.org/10.1103/RevModPhys.79.1217}
  {\bibfield  {journal} {\bibinfo  {journal} {Rev. Mod. Phys.}\ }\textbf
  {\bibinfo {volume} {79}},\ \bibinfo {pages} {1217} (\bibinfo {year}
  {2007})}\BibitemShut {NoStop}%
\bibitem [{\citenamefont {Burkard}\ \emph {et~al.}(2023)\citenamefont
  {Burkard}, \citenamefont {Ladd}, \citenamefont {Pan}, \citenamefont
  {Nichol},\ and\ \citenamefont {Petta}}]{burkard_semiconductor_2023}%
  \BibitemOpen
  \bibfield  {author} {\bibinfo {author} {\bibfnamefont {G.}~\bibnamefont
  {Burkard}}, \bibinfo {author} {\bibfnamefont {T.~D.}\ \bibnamefont {Ladd}},
  \bibinfo {author} {\bibfnamefont {A.}~\bibnamefont {Pan}}, \bibinfo {author}
  {\bibfnamefont {J.~M.}\ \bibnamefont {Nichol}},\ and\ \bibinfo {author}
  {\bibfnamefont {J.~R.}\ \bibnamefont {Petta}},\ }\bibfield  {title} {\bibinfo
  {title} {Semiconductor spin qubits},\ }\href@noop {} {\bibfield  {journal}
  {\bibinfo  {journal} {Reviews of Modern Physics}\ }\textbf {\bibinfo {volume}
  {95}},\ \bibinfo {pages} {025003} (\bibinfo {year} {2023})}\BibitemShut
  {NoStop}%
\bibitem [{\citenamefont {Hansen}\ \emph {et~al.}(2022)\citenamefont {Hansen},
  \citenamefont {Seedhouse}, \citenamefont {Chan}, \citenamefont {Hudson},
  \citenamefont {Itoh}, \citenamefont {Laucht}, \citenamefont {Saraiva},
  \citenamefont {Yang},\ and\ \citenamefont
  {Dzurak}}]{hansen_implementation_2022}%
  \BibitemOpen
  \bibfield  {author} {\bibinfo {author} {\bibfnamefont {I.}~\bibnamefont
  {Hansen}}, \bibinfo {author} {\bibfnamefont {A.~E.}\ \bibnamefont
  {Seedhouse}}, \bibinfo {author} {\bibfnamefont {K.~W.}\ \bibnamefont {Chan}},
  \bibinfo {author} {\bibfnamefont {F.~E.}\ \bibnamefont {Hudson}}, \bibinfo
  {author} {\bibfnamefont {K.~M.}\ \bibnamefont {Itoh}}, \bibinfo {author}
  {\bibfnamefont {A.}~\bibnamefont {Laucht}}, \bibinfo {author} {\bibfnamefont
  {A.}~\bibnamefont {Saraiva}}, \bibinfo {author} {\bibfnamefont {C.~H.}\
  \bibnamefont {Yang}},\ and\ \bibinfo {author} {\bibfnamefont {A.~S.}\
  \bibnamefont {Dzurak}},\ }\bibfield  {title} {\bibinfo {title}
  {Implementation of an advanced dressing protocol for global qubit control in
  silicon},\ }\href {https://doi.org/10.1063/5.0096467} {\bibfield  {journal}
  {\bibinfo  {journal} {Applied Physics Reviews}\ }\textbf {\bibinfo {volume}
  {9}},\ \bibinfo {pages} {031409} (\bibinfo {year} {2022})}\BibitemShut
  {NoStop}%
\bibitem [{\citenamefont {Veldhorst}\ \emph {et~al.}(2014)\citenamefont
  {Veldhorst}, \citenamefont {Hwang}, \citenamefont {Yang}, \citenamefont
  {Leenstra}, \citenamefont {de~Ronde}, \citenamefont {Dehollain},
  \citenamefont {Muhonen}, \citenamefont {Hudson}, \citenamefont {Itoh},
  \citenamefont {Morello},\ and\ \citenamefont
  {Dzurak}}]{veldhorst_addressable_2014}%
  \BibitemOpen
  \bibfield  {author} {\bibinfo {author} {\bibfnamefont {M.}~\bibnamefont
  {Veldhorst}}, \bibinfo {author} {\bibfnamefont {J.~C.~C.}\ \bibnamefont
  {Hwang}}, \bibinfo {author} {\bibfnamefont {C.~H.}\ \bibnamefont {Yang}},
  \bibinfo {author} {\bibfnamefont {A.~W.}\ \bibnamefont {Leenstra}}, \bibinfo
  {author} {\bibfnamefont {B.}~\bibnamefont {de~Ronde}}, \bibinfo {author}
  {\bibfnamefont {J.~P.}\ \bibnamefont {Dehollain}}, \bibinfo {author}
  {\bibfnamefont {J.~T.}\ \bibnamefont {Muhonen}}, \bibinfo {author}
  {\bibfnamefont {F.~E.}\ \bibnamefont {Hudson}}, \bibinfo {author}
  {\bibfnamefont {K.~M.}\ \bibnamefont {Itoh}}, \bibinfo {author}
  {\bibfnamefont {A.}~\bibnamefont {Morello}},\ and\ \bibinfo {author}
  {\bibfnamefont {A.~S.}\ \bibnamefont {Dzurak}},\ }\bibfield  {title}
  {\bibinfo {title} {An addressable quantum dot qubit with fault-tolerant
  control-fidelity},\ }\href {https://doi.org/10.1038/nnano.2014.216}
  {\bibfield  {journal} {\bibinfo  {journal} {Nature Nanotechnology}\ }\textbf
  {\bibinfo {volume} {9}},\ \bibinfo {pages} {981} (\bibinfo {year} {2014})},\
  \bibinfo {note} {publisher: Nature Publishing Group}\BibitemShut {NoStop}%
\bibitem [{\citenamefont {Mills}\ \emph
  {et~al.}(2022{\natexlab{a}})\citenamefont {Mills}, \citenamefont {Guinn},
  \citenamefont {Feldman}, \citenamefont {Sigillito}, \citenamefont {Gullans},
  \citenamefont {Rakher}, \citenamefont {Kerckhoff}, \citenamefont {Jackson},\
  and\ \citenamefont {Petta}}]{mills_high-fidelity_2022}%
  \BibitemOpen
  \bibfield  {author} {\bibinfo {author} {\bibfnamefont {A.~R.}\ \bibnamefont
  {Mills}}, \bibinfo {author} {\bibfnamefont {C.~R.}\ \bibnamefont {Guinn}},
  \bibinfo {author} {\bibfnamefont {M.~M.}\ \bibnamefont {Feldman}}, \bibinfo
  {author} {\bibfnamefont {A.~J.}\ \bibnamefont {Sigillito}}, \bibinfo {author}
  {\bibfnamefont {M.~J.}\ \bibnamefont {Gullans}}, \bibinfo {author}
  {\bibfnamefont {M.~T.}\ \bibnamefont {Rakher}}, \bibinfo {author}
  {\bibfnamefont {J.}~\bibnamefont {Kerckhoff}}, \bibinfo {author}
  {\bibfnamefont {C.~A.~C.}\ \bibnamefont {Jackson}},\ and\ \bibinfo {author}
  {\bibfnamefont {J.~R.}\ \bibnamefont {Petta}},\ }\bibfield  {title} {\bibinfo
  {title} {High-{Fidelity} {State} {Preparation}, {Quantum} {Control}, and
  {Readout} of an {Isotopically} {Enriched} {Silicon} {Spin} {Qubit}},\ }\href
  {https://doi.org/10.1103/PhysRevApplied.18.064028} {\bibfield  {journal}
  {\bibinfo  {journal} {Physical Review Applied}\ }\textbf {\bibinfo {volume}
  {18}},\ \bibinfo {pages} {064028} (\bibinfo {year} {2022}{\natexlab{a}})},\
  \bibinfo {note} {publisher: American Physical Society}\BibitemShut {NoStop}%
\bibitem [{\citenamefont {Takeda}\ \emph {et~al.}(2016)\citenamefont {Takeda},
  \citenamefont {Kamioka}, \citenamefont {Otsuka}, \citenamefont {Yoneda},
  \citenamefont {Nakajima}, \citenamefont {Delbecq}, \citenamefont {Amaha},
  \citenamefont {Allison}, \citenamefont {Kodera}, \citenamefont {Oda},\ and\
  \citenamefont {Tarucha}}]{takeda_fault-tolerant_2016}%
  \BibitemOpen
  \bibfield  {author} {\bibinfo {author} {\bibfnamefont {K.}~\bibnamefont
  {Takeda}}, \bibinfo {author} {\bibfnamefont {J.}~\bibnamefont {Kamioka}},
  \bibinfo {author} {\bibfnamefont {T.}~\bibnamefont {Otsuka}}, \bibinfo
  {author} {\bibfnamefont {J.}~\bibnamefont {Yoneda}}, \bibinfo {author}
  {\bibfnamefont {T.}~\bibnamefont {Nakajima}}, \bibinfo {author}
  {\bibfnamefont {M.~R.}\ \bibnamefont {Delbecq}}, \bibinfo {author}
  {\bibfnamefont {S.}~\bibnamefont {Amaha}}, \bibinfo {author} {\bibfnamefont
  {G.}~\bibnamefont {Allison}}, \bibinfo {author} {\bibfnamefont
  {T.}~\bibnamefont {Kodera}}, \bibinfo {author} {\bibfnamefont
  {S.}~\bibnamefont {Oda}},\ and\ \bibinfo {author} {\bibfnamefont
  {S.}~\bibnamefont {Tarucha}},\ }\bibfield  {title} {\bibinfo {title} {A
  fault-tolerant addressable spin qubit in a natural silicon quantum dot},\
  }\href {https://doi.org/10.1126/sciadv.1600694} {\bibfield  {journal}
  {\bibinfo  {journal} {Science Advances}\ }\textbf {\bibinfo {volume} {2}},\
  \bibinfo {pages} {e1600694} (\bibinfo {year} {2016})},\ \bibinfo {note}
  {publisher: American Association for the Advancement of Science}\BibitemShut
  {NoStop}%
\bibitem [{\citenamefont {Mills}\ \emph
  {et~al.}(2022{\natexlab{b}})\citenamefont {Mills}, \citenamefont {Guinn},
  \citenamefont {Gullans}, \citenamefont {Sigillito}, \citenamefont {Feldman},
  \citenamefont {Nielsen},\ and\ \citenamefont {Petta}}]{mills_two-qubit_2022}%
  \BibitemOpen
  \bibfield  {author} {\bibinfo {author} {\bibfnamefont {A.~R.}\ \bibnamefont
  {Mills}}, \bibinfo {author} {\bibfnamefont {C.~R.}\ \bibnamefont {Guinn}},
  \bibinfo {author} {\bibfnamefont {M.~J.}\ \bibnamefont {Gullans}}, \bibinfo
  {author} {\bibfnamefont {A.~J.}\ \bibnamefont {Sigillito}}, \bibinfo {author}
  {\bibfnamefont {M.~M.}\ \bibnamefont {Feldman}}, \bibinfo {author}
  {\bibfnamefont {E.}~\bibnamefont {Nielsen}},\ and\ \bibinfo {author}
  {\bibfnamefont {J.~R.}\ \bibnamefont {Petta}},\ }\bibfield  {title} {\bibinfo
  {title} {Two-qubit silicon quantum processor with operation fidelity
  exceeding 99\%},\ }\href {https://doi.org/10.1126/sciadv.abn5130} {\bibfield
  {journal} {\bibinfo  {journal} {Science Advances}\ }\textbf {\bibinfo
  {volume} {8}},\ \bibinfo {pages} {eabn5130} (\bibinfo {year}
  {2022}{\natexlab{b}})},\ \bibinfo {note} {publisher: American Association for
  the Advancement of Science}\BibitemShut {NoStop}%
\bibitem [{\citenamefont {Croot}\ \emph {et~al.}(2020)\citenamefont {Croot},
  \citenamefont {Mi}, \citenamefont {Putz}, \citenamefont {Benito},
  \citenamefont {Borjans}, \citenamefont {Burkard},\ and\ \citenamefont
  {Petta}}]{croot_flopping-mode_2020}%
  \BibitemOpen
  \bibfield  {author} {\bibinfo {author} {\bibfnamefont {X.}~\bibnamefont
  {Croot}}, \bibinfo {author} {\bibfnamefont {X.}~\bibnamefont {Mi}}, \bibinfo
  {author} {\bibfnamefont {S.}~\bibnamefont {Putz}}, \bibinfo {author}
  {\bibfnamefont {M.}~\bibnamefont {Benito}}, \bibinfo {author} {\bibfnamefont
  {F.}~\bibnamefont {Borjans}}, \bibinfo {author} {\bibfnamefont
  {G.}~\bibnamefont {Burkard}},\ and\ \bibinfo {author} {\bibfnamefont {J.~R.}\
  \bibnamefont {Petta}},\ }\bibfield  {title} {\bibinfo {title} {Flopping-mode
  electric dipole spin resonance},\ }\href
  {https://doi.org/10.1103/PhysRevResearch.2.012006} {\bibfield  {journal}
  {\bibinfo  {journal} {Physical Review Research}\ }\textbf {\bibinfo {volume}
  {2}},\ \bibinfo {pages} {012006(R)} (\bibinfo {year} {2020})},\ \bibinfo
  {note} {publisher: American Physical Society}\BibitemShut {NoStop}%
\bibitem [{\citenamefont {Weinstein}\ \emph {et~al.}(2023)\citenamefont
  {Weinstein}, \citenamefont {Reed}, \citenamefont {Jones}, \citenamefont
  {Andrews}, \citenamefont {Barnes}, \citenamefont {Blumoff}, \citenamefont
  {Euliss}, \citenamefont {Eng}, \citenamefont {Fong}, \citenamefont {Ha},
  \citenamefont {Hulbert}, \citenamefont {Jackson}, \citenamefont {Jura},
  \citenamefont {Keating}, \citenamefont {Kerckhoff}, \citenamefont {Kiselev},
  \citenamefont {Matten}, \citenamefont {Sabbir}, \citenamefont {Smith},
  \citenamefont {Wright}, \citenamefont {Rakher}, \citenamefont {Ladd},\ and\
  \citenamefont {Borselli}}]{weinstein_universal_2023}%
  \BibitemOpen
  \bibfield  {author} {\bibinfo {author} {\bibfnamefont {A.}~\bibnamefont
  {Weinstein}}, \bibinfo {author} {\bibfnamefont {M.}~\bibnamefont {Reed}},
  \bibinfo {author} {\bibfnamefont {A.}~\bibnamefont {Jones}}, \bibinfo
  {author} {\bibfnamefont {R.}~\bibnamefont {Andrews}}, \bibinfo {author}
  {\bibfnamefont {D.}~\bibnamefont {Barnes}}, \bibinfo {author} {\bibfnamefont
  {J.}~\bibnamefont {Blumoff}}, \bibinfo {author} {\bibfnamefont
  {L.}~\bibnamefont {Euliss}}, \bibinfo {author} {\bibfnamefont
  {K.}~\bibnamefont {Eng}}, \bibinfo {author} {\bibfnamefont {B.}~\bibnamefont
  {Fong}}, \bibinfo {author} {\bibfnamefont {S.}~\bibnamefont {Ha}}, \bibinfo
  {author} {\bibfnamefont {D.}~\bibnamefont {Hulbert}}, \bibinfo {author}
  {\bibfnamefont {C.}~\bibnamefont {Jackson}}, \bibinfo {author} {\bibfnamefont
  {M.}~\bibnamefont {Jura}}, \bibinfo {author} {\bibfnamefont {T.}~\bibnamefont
  {Keating}}, \bibinfo {author} {\bibfnamefont {J.}~\bibnamefont {Kerckhoff}},
  \bibinfo {author} {\bibfnamefont {A.}~\bibnamefont {Kiselev}}, \bibinfo
  {author} {\bibfnamefont {J.}~\bibnamefont {Matten}}, \bibinfo {author}
  {\bibfnamefont {G.}~\bibnamefont {Sabbir}}, \bibinfo {author} {\bibfnamefont
  {A.}~\bibnamefont {Smith}}, \bibinfo {author} {\bibfnamefont
  {J.}~\bibnamefont {Wright}}, \bibinfo {author} {\bibfnamefont
  {M.}~\bibnamefont {Rakher}}, \bibinfo {author} {\bibfnamefont
  {T.}~\bibnamefont {Ladd}},\ and\ \bibinfo {author} {\bibfnamefont
  {M.}~\bibnamefont {Borselli}},\ }\bibfield  {title} {\bibinfo {title}
  {Universal logic with encoded spin qubits in silicon.},\ }\href@noop {}
  {\bibfield  {journal} {\bibinfo  {journal} {Nature}\ }\textbf {\bibinfo
  {volume} {615}},\ \bibinfo {pages} {817} (\bibinfo {year}
  {2023})}\BibitemShut {NoStop}%
\bibitem [{\citenamefont {Neyens}\ \emph {et~al.}(2024)\citenamefont {Neyens},
  \citenamefont {Zietz}, \citenamefont {Watson}, \citenamefont {Luthi},
  \citenamefont {Nethwewala}, \citenamefont {George}, \citenamefont {Henry},
  \citenamefont {Islam}, \citenamefont {Wagner}, \citenamefont {Borjans},
  \citenamefont {Connors}, \citenamefont {Corrigan}, \citenamefont {Curry},
  \citenamefont {Keith}, \citenamefont {Kotlyar}, \citenamefont {Lampert},
  \citenamefont {Mądzik}, \citenamefont {Millard}, \citenamefont {Mohiyaddin},
  \citenamefont {Pellerano}, \citenamefont {Pillarisetty}, \citenamefont
  {Ramsey}, \citenamefont {Savytskyy}, \citenamefont {Schaal}, \citenamefont
  {Zheng}, \citenamefont {Ziegler}, \citenamefont {Bishop}, \citenamefont
  {Bojarski}, \citenamefont {Roberts},\ and\ \citenamefont
  {Clarke}}]{neyens_probing_2024}%
  \BibitemOpen
  \bibfield  {author} {\bibinfo {author} {\bibfnamefont {S.}~\bibnamefont
  {Neyens}}, \bibinfo {author} {\bibfnamefont {O.}~\bibnamefont {Zietz}},
  \bibinfo {author} {\bibfnamefont {T.}~\bibnamefont {Watson}}, \bibinfo
  {author} {\bibfnamefont {F.}~\bibnamefont {Luthi}}, \bibinfo {author}
  {\bibfnamefont {A.}~\bibnamefont {Nethwewala}}, \bibinfo {author}
  {\bibfnamefont {H.}~\bibnamefont {George}}, \bibinfo {author} {\bibfnamefont
  {E.}~\bibnamefont {Henry}}, \bibinfo {author} {\bibfnamefont
  {M.}~\bibnamefont {Islam}}, \bibinfo {author} {\bibfnamefont
  {A.}~\bibnamefont {Wagner}}, \bibinfo {author} {\bibfnamefont
  {F.}~\bibnamefont {Borjans}}, \bibinfo {author} {\bibfnamefont
  {E.}~\bibnamefont {Connors}}, \bibinfo {author} {\bibfnamefont
  {J.}~\bibnamefont {Corrigan}}, \bibinfo {author} {\bibfnamefont
  {M.}~\bibnamefont {Curry}}, \bibinfo {author} {\bibfnamefont
  {D.}~\bibnamefont {Keith}}, \bibinfo {author} {\bibfnamefont
  {R.}~\bibnamefont {Kotlyar}}, \bibinfo {author} {\bibfnamefont
  {L.}~\bibnamefont {Lampert}}, \bibinfo {author} {\bibfnamefont
  {M.}~\bibnamefont {Mądzik}}, \bibinfo {author} {\bibfnamefont
  {K.}~\bibnamefont {Millard}}, \bibinfo {author} {\bibfnamefont
  {F.}~\bibnamefont {Mohiyaddin}}, \bibinfo {author} {\bibfnamefont
  {S.}~\bibnamefont {Pellerano}}, \bibinfo {author} {\bibfnamefont
  {R.}~\bibnamefont {Pillarisetty}}, \bibinfo {author} {\bibfnamefont
  {M.}~\bibnamefont {Ramsey}}, \bibinfo {author} {\bibfnamefont
  {R.}~\bibnamefont {Savytskyy}}, \bibinfo {author} {\bibfnamefont
  {S.}~\bibnamefont {Schaal}}, \bibinfo {author} {\bibfnamefont
  {G.}~\bibnamefont {Zheng}}, \bibinfo {author} {\bibfnamefont
  {J.}~\bibnamefont {Ziegler}}, \bibinfo {author} {\bibfnamefont
  {N.}~\bibnamefont {Bishop}}, \bibinfo {author} {\bibfnamefont
  {S.}~\bibnamefont {Bojarski}}, \bibinfo {author} {\bibfnamefont
  {J.}~\bibnamefont {Roberts}},\ and\ \bibinfo {author} {\bibfnamefont
  {J.}~\bibnamefont {Clarke}},\ }\bibfield  {title} {\bibinfo {title} {Probing
  single electrons across 300-mm spin qubit wafers.},\ }\href@noop {}
  {\bibfield  {journal} {\bibinfo  {journal} {Nature}\ }\textbf {\bibinfo
  {volume} {629}},\ \bibinfo {pages} {80} (\bibinfo {year} {2024})}\BibitemShut
  {NoStop}%
\bibitem [{\citenamefont {Elsayed}\ \emph {et~al.}(2022)\citenamefont
  {Elsayed}, \citenamefont {Shehata}, \citenamefont {Godfrin}, \citenamefont
  {Kubicek}, \citenamefont {Massar}, \citenamefont {Canvel}, \citenamefont
  {Jussot}, \citenamefont {Simion}, \citenamefont {Mongillo},\ and\
  \citenamefont {Wan}}]{elsayed_low_2022}%
  \BibitemOpen
  \bibfield  {author} {\bibinfo {author} {\bibfnamefont {A.}~\bibnamefont
  {Elsayed}}, \bibinfo {author} {\bibfnamefont {M.}~\bibnamefont {Shehata}},
  \bibinfo {author} {\bibfnamefont {C.}~\bibnamefont {Godfrin}}, \bibinfo
  {author} {\bibfnamefont {S.}~\bibnamefont {Kubicek}}, \bibinfo {author}
  {\bibfnamefont {S.}~\bibnamefont {Massar}}, \bibinfo {author} {\bibfnamefont
  {Y.}~\bibnamefont {Canvel}}, \bibinfo {author} {\bibfnamefont
  {J.}~\bibnamefont {Jussot}}, \bibinfo {author} {\bibfnamefont
  {G.}~\bibnamefont {Simion}}, \bibinfo {author} {\bibfnamefont
  {M.}~\bibnamefont {Mongillo}},\ and\ \bibinfo {author} {\bibfnamefont
  {D.}~\bibnamefont {Wan}},\ }\href@noop {} {\bibinfo {title} {Low charge noise
  quantum dots with industrial {CMOS} manufacturing}} (\bibinfo {year}
  {2022})\BibitemShut {NoStop}%
\bibitem [{\citenamefont {Ha}\ \emph {et~al.}(2022)\citenamefont {Ha},
  \citenamefont {Ha}, \citenamefont {Choi}, \citenamefont {Tang}, \citenamefont
  {Schmitz}, \citenamefont {Levendorf}, \citenamefont {Lee}, \citenamefont
  {Chappell}, \citenamefont {Adams}, \citenamefont {Hulbert}, \citenamefont
  {Acuna}, \citenamefont {Noah}, \citenamefont {Matten}, \citenamefont {Jura},
  \citenamefont {Wright}, \citenamefont {Rakher},\ and\ \citenamefont
  {Borselli}}]{ha_flexible_2022}%
  \BibitemOpen
  \bibfield  {author} {\bibinfo {author} {\bibfnamefont {W.}~\bibnamefont
  {Ha}}, \bibinfo {author} {\bibfnamefont {S.~D.}\ \bibnamefont {Ha}}, \bibinfo
  {author} {\bibfnamefont {M.~D.}\ \bibnamefont {Choi}}, \bibinfo {author}
  {\bibfnamefont {Y.}~\bibnamefont {Tang}}, \bibinfo {author} {\bibfnamefont
  {A.~E.}\ \bibnamefont {Schmitz}}, \bibinfo {author} {\bibfnamefont {M.~P.}\
  \bibnamefont {Levendorf}}, \bibinfo {author} {\bibfnamefont {K.}~\bibnamefont
  {Lee}}, \bibinfo {author} {\bibfnamefont {J.~M.}\ \bibnamefont {Chappell}},
  \bibinfo {author} {\bibfnamefont {T.~S.}\ \bibnamefont {Adams}}, \bibinfo
  {author} {\bibfnamefont {D.~R.}\ \bibnamefont {Hulbert}}, \bibinfo {author}
  {\bibfnamefont {E.}~\bibnamefont {Acuna}}, \bibinfo {author} {\bibfnamefont
  {R.~S.}\ \bibnamefont {Noah}}, \bibinfo {author} {\bibfnamefont {J.~W.}\
  \bibnamefont {Matten}}, \bibinfo {author} {\bibfnamefont {M.~P.}\
  \bibnamefont {Jura}}, \bibinfo {author} {\bibfnamefont {J.~A.}\ \bibnamefont
  {Wright}}, \bibinfo {author} {\bibfnamefont {M.~T.}\ \bibnamefont {Rakher}},\
  and\ \bibinfo {author} {\bibfnamefont {M.~G.}\ \bibnamefont {Borselli}},\
  }\bibfield  {title} {\bibinfo {title} {A {Flexible} {Design} {Platform} for
  {Si}/{SiGe} {Exchange}-{Only} {Qubits} with {Low} {Disorder}},\ }\href
  {https://doi.org/10.1021/acs.nanolett.1c03026} {\bibfield  {journal}
  {\bibinfo  {journal} {Nano Letters}\ }\textbf {\bibinfo {volume} {22}},\
  \bibinfo {pages} {1443} (\bibinfo {year} {2022})},\ \bibinfo {note}
  {publisher: American Chemical Society}\BibitemShut {NoStop}%
\bibitem [{\citenamefont {Fowler}\ \emph {et~al.}(2012)\citenamefont {Fowler},
  \citenamefont {Mariantoni}, \citenamefont {Martinis},\ and\ \citenamefont
  {Cleland}}]{fowler_surface_2012}%
  \BibitemOpen
  \bibfield  {author} {\bibinfo {author} {\bibfnamefont {A.~G.}\ \bibnamefont
  {Fowler}}, \bibinfo {author} {\bibfnamefont {M.}~\bibnamefont {Mariantoni}},
  \bibinfo {author} {\bibfnamefont {J.~M.}\ \bibnamefont {Martinis}},\ and\
  \bibinfo {author} {\bibfnamefont {A.~N.}\ \bibnamefont {Cleland}},\
  }\bibfield  {title} {\bibinfo {title} {Surface codes: {Towards} practical
  large-scale quantum computation},\ }\href
  {https://doi.org/10.1103/PhysRevA.86.032324} {\bibfield  {journal} {\bibinfo
  {journal} {Physical Review A}\ }\textbf {\bibinfo {volume} {86}},\ \bibinfo
  {pages} {032324} (\bibinfo {year} {2012})}\BibitemShut {NoStop}%
\bibitem [{\citenamefont {Vandersypen}\ \emph {et~al.}(2017)\citenamefont
  {Vandersypen}, \citenamefont {Bluhm}, \citenamefont {Clarke}, \citenamefont
  {Dzurak}, \citenamefont {Ishihara}, \citenamefont {Morello}, \citenamefont
  {Reilly}, \citenamefont {Schreiber},\ and\ \citenamefont
  {Veldhorst}}]{vandersypen_interfacing_2017}%
  \BibitemOpen
  \bibfield  {author} {\bibinfo {author} {\bibfnamefont {L.~M.~K.}\
  \bibnamefont {Vandersypen}}, \bibinfo {author} {\bibfnamefont
  {H.}~\bibnamefont {Bluhm}}, \bibinfo {author} {\bibfnamefont {J.~S.}\
  \bibnamefont {Clarke}}, \bibinfo {author} {\bibfnamefont {A.~S.}\
  \bibnamefont {Dzurak}}, \bibinfo {author} {\bibfnamefont {R.}~\bibnamefont
  {Ishihara}}, \bibinfo {author} {\bibfnamefont {A.}~\bibnamefont {Morello}},
  \bibinfo {author} {\bibfnamefont {D.~J.}\ \bibnamefont {Reilly}}, \bibinfo
  {author} {\bibfnamefont {L.~R.}\ \bibnamefont {Schreiber}},\ and\ \bibinfo
  {author} {\bibfnamefont {M.}~\bibnamefont {Veldhorst}},\ }\bibfield  {title}
  {\bibinfo {title} {Interfacing spin qubits in quantum dots and donors - hot,
  dense and coherent},\ }\href {https://doi.org/10.1038/s41534-017-0038-y}
  {\bibfield  {journal} {\bibinfo  {journal} {npj Quantum Information}\
  }\textbf {\bibinfo {volume} {3}},\ \bibinfo {pages} {34} (\bibinfo {year}
  {2017})},\ \bibinfo {note} {arXiv:1612.05936 [cond-mat,
  physics:quant-ph]}\BibitemShut {NoStop}%
\bibitem [{\citenamefont {Koppens}\ \emph {et~al.}(2008)\citenamefont
  {Koppens}, \citenamefont {Nowack},\ and\ \citenamefont
  {Vandersypen}}]{koppens_spin_2008}%
  \BibitemOpen
  \bibfield  {author} {\bibinfo {author} {\bibfnamefont {F.~H.~L.}\
  \bibnamefont {Koppens}}, \bibinfo {author} {\bibfnamefont {K.~C.}\
  \bibnamefont {Nowack}},\ and\ \bibinfo {author} {\bibfnamefont {L.~M.~K.}\
  \bibnamefont {Vandersypen}},\ }\bibfield  {title} {\bibinfo {title} {Spin
  {Echo} of a {Single} {Electron} {Spin} in a {Quantum} {Dot}},\ }\href
  {https://doi.org/10.1103/PhysRevLett.100.236802} {\bibfield  {journal}
  {\bibinfo  {journal} {Physical Review Letters}\ }\textbf {\bibinfo {volume}
  {100}},\ \bibinfo {pages} {236802} (\bibinfo {year} {2008})},\ \bibinfo
  {note} {publisher: American Physical Society}\BibitemShut {NoStop}%
\bibitem [{\citenamefont {Golovach}\ \emph {et~al.}(2006)\citenamefont
  {Golovach}, \citenamefont {Borhani},\ and\ \citenamefont
  {Loss}}]{golovach_electric-dipole-induced_2006}%
  \BibitemOpen
  \bibfield  {author} {\bibinfo {author} {\bibfnamefont {V.~N.}\ \bibnamefont
  {Golovach}}, \bibinfo {author} {\bibfnamefont {M.}~\bibnamefont {Borhani}},\
  and\ \bibinfo {author} {\bibfnamefont {D.}~\bibnamefont {Loss}},\ }\bibfield
  {title} {\bibinfo {title} {Electric-dipole-induced spin resonance in quantum
  dots},\ }\href {https://doi.org/10.1103/PhysRevB.74.165319} {\bibfield
  {journal} {\bibinfo  {journal} {Physical Review B}\ }\textbf {\bibinfo
  {volume} {74}},\ \bibinfo {pages} {165319} (\bibinfo {year} {2006})},\
  \bibinfo {note} {publisher: American Physical Society}\BibitemShut {NoStop}%
\bibitem [{\citenamefont {Petta}\ \emph {et~al.}(2005)\citenamefont {Petta},
  \citenamefont {Johnson}, \citenamefont {Taylor}, \citenamefont {Laird},
  \citenamefont {Yacoby}, \citenamefont {Lukin}, \citenamefont {Marcus},
  \citenamefont {Hanson},\ and\ \citenamefont {Gossard}}]{petta_coherent_2005}%
  \BibitemOpen
  \bibfield  {author} {\bibinfo {author} {\bibfnamefont {J.~R.}\ \bibnamefont
  {Petta}}, \bibinfo {author} {\bibfnamefont {A.~C.}\ \bibnamefont {Johnson}},
  \bibinfo {author} {\bibfnamefont {J.~M.}\ \bibnamefont {Taylor}}, \bibinfo
  {author} {\bibfnamefont {E.~A.}\ \bibnamefont {Laird}}, \bibinfo {author}
  {\bibfnamefont {A.}~\bibnamefont {Yacoby}}, \bibinfo {author} {\bibfnamefont
  {M.~D.}\ \bibnamefont {Lukin}}, \bibinfo {author} {\bibfnamefont {C.~M.}\
  \bibnamefont {Marcus}}, \bibinfo {author} {\bibfnamefont {M.~P.}\
  \bibnamefont {Hanson}},\ and\ \bibinfo {author} {\bibfnamefont {A.~C.}\
  \bibnamefont {Gossard}},\ }\bibfield  {title} {\bibinfo {title} {Coherent
  {Manipulation} of {Coupled} {Electron} {Spins} in {Semiconductor} {Quantum}
  {Dots}},\ }\href {https://doi.org/10.1126/science.1116955} {\bibfield
  {journal} {\bibinfo  {journal} {Science}\ }\textbf {\bibinfo {volume}
  {309}},\ \bibinfo {pages} {2180} (\bibinfo {year} {2005})},\ \bibinfo {note}
  {publisher: American Association for the Advancement of Science}\BibitemShut
  {NoStop}%
\bibitem [{\citenamefont {Loss}\ and\ \citenamefont
  {DiVincenzo}(1998)}]{loss_quantum_1998}%
  \BibitemOpen
  \bibfield  {author} {\bibinfo {author} {\bibfnamefont {D.}~\bibnamefont
  {Loss}}\ and\ \bibinfo {author} {\bibfnamefont {D.~P.}\ \bibnamefont
  {DiVincenzo}},\ }\bibfield  {title} {\bibinfo {title} {Quantum computation
  with quantum dots},\ }\href {https://doi.org/10.1103/PhysRevA.57.120}
  {\bibfield  {journal} {\bibinfo  {journal} {Physical Review A}\ }\textbf
  {\bibinfo {volume} {57}},\ \bibinfo {pages} {120} (\bibinfo {year} {1998})},\
  \bibinfo {note} {publisher: American Physical Society}\BibitemShut {NoStop}%
\bibitem [{\citenamefont {Hu}\ and\ \citenamefont
  {Das~Sarma}(2000)}]{hu_hilbert-space_2000}%
  \BibitemOpen
  \bibfield  {author} {\bibinfo {author} {\bibfnamefont {X.}~\bibnamefont
  {Hu}}\ and\ \bibinfo {author} {\bibfnamefont {S.}~\bibnamefont {Das~Sarma}},\
  }\bibfield  {title} {\bibinfo {title} {Hilbert-space structure of a
  solid-state quantum computer: {Two}-electron states of a double-quantum-dot
  artificial molecule},\ }\href {https://doi.org/10.1103/PhysRevA.61.062301}
  {\bibfield  {journal} {\bibinfo  {journal} {Physical Review A}\ }\textbf
  {\bibinfo {volume} {61}},\ \bibinfo {pages} {062301} (\bibinfo {year}
  {2000})},\ \bibinfo {note} {publisher: American Physical Society}\BibitemShut
  {NoStop}%
\bibitem [{\citenamefont {Burkard}\ \emph {et~al.}(1999)\citenamefont
  {Burkard}, \citenamefont {Loss},\ and\ \citenamefont
  {DiVincenzo}}]{burkard_coupled_1999}%
  \BibitemOpen
  \bibfield  {author} {\bibinfo {author} {\bibfnamefont {G.}~\bibnamefont
  {Burkard}}, \bibinfo {author} {\bibfnamefont {D.}~\bibnamefont {Loss}},\ and\
  \bibinfo {author} {\bibfnamefont {D.~P.}\ \bibnamefont {DiVincenzo}},\
  }\bibfield  {title} {\bibinfo {title} {Coupled quantum dots as quantum
  gates},\ }\href {https://doi.org/10.1103/PhysRevB.59.2070} {\bibfield
  {journal} {\bibinfo  {journal} {Physical Review B}\ }\textbf {\bibinfo
  {volume} {59}},\ \bibinfo {pages} {2070} (\bibinfo {year} {1999})},\ \bibinfo
  {note} {publisher: American Physical Society}\BibitemShut {NoStop}%
\bibitem [{\citenamefont {Medford}\ \emph {et~al.}(2013)\citenamefont
  {Medford}, \citenamefont {Beil}, \citenamefont {Taylor}, \citenamefont
  {Bartlett}, \citenamefont {Doherty}, \citenamefont {Rashba}, \citenamefont
  {DiVincenzo}, \citenamefont {Lu}, \citenamefont {Gossard},\ and\
  \citenamefont {Marcus}}]{medford_self-consistent_2013}%
  \BibitemOpen
  \bibfield  {author} {\bibinfo {author} {\bibfnamefont {J.}~\bibnamefont
  {Medford}}, \bibinfo {author} {\bibfnamefont {J.}~\bibnamefont {Beil}},
  \bibinfo {author} {\bibfnamefont {J.~M.}\ \bibnamefont {Taylor}}, \bibinfo
  {author} {\bibfnamefont {S.~D.}\ \bibnamefont {Bartlett}}, \bibinfo {author}
  {\bibfnamefont {A.~C.}\ \bibnamefont {Doherty}}, \bibinfo {author}
  {\bibfnamefont {E.~I.}\ \bibnamefont {Rashba}}, \bibinfo {author}
  {\bibfnamefont {D.~P.}\ \bibnamefont {DiVincenzo}}, \bibinfo {author}
  {\bibfnamefont {H.}~\bibnamefont {Lu}}, \bibinfo {author} {\bibfnamefont
  {A.~C.}\ \bibnamefont {Gossard}},\ and\ \bibinfo {author} {\bibfnamefont
  {C.~M.}\ \bibnamefont {Marcus}},\ }\bibfield  {title} {\bibinfo {title}
  {Self-{Consistent} {Measurement} and {State} {Tomography} of an
  {Exchange}-{Only} {Spin} {Qubit}},\ }\href
  {https://doi.org/10.1038/nnano.2013.168} {\bibfield  {journal} {\bibinfo
  {journal} {Nature Nanotechnology}\ }\textbf {\bibinfo {volume} {8}},\
  \bibinfo {pages} {654} (\bibinfo {year} {2013})},\ \bibinfo {note}
  {arXiv:1302.1933 [cond-mat, physics:quant-ph]}\BibitemShut {NoStop}%
\bibitem [{\citenamefont {Benito}\ \emph {et~al.}(2019)\citenamefont {Benito},
  \citenamefont {Croot}, \citenamefont {Adelsberger}, \citenamefont {Putz},
  \citenamefont {Mi}, \citenamefont {Petta},\ and\ \citenamefont
  {Burkard}}]{benito_electric-field_2019}%
  \BibitemOpen
  \bibfield  {author} {\bibinfo {author} {\bibfnamefont {M.}~\bibnamefont
  {Benito}}, \bibinfo {author} {\bibfnamefont {X.}~\bibnamefont {Croot}},
  \bibinfo {author} {\bibfnamefont {C.}~\bibnamefont {Adelsberger}}, \bibinfo
  {author} {\bibfnamefont {S.}~\bibnamefont {Putz}}, \bibinfo {author}
  {\bibfnamefont {X.}~\bibnamefont {Mi}}, \bibinfo {author} {\bibfnamefont
  {J.~R.}\ \bibnamefont {Petta}},\ and\ \bibinfo {author} {\bibfnamefont
  {G.}~\bibnamefont {Burkard}},\ }\bibfield  {title} {\bibinfo {title}
  {Electric-field control and noise protection of the flopping-mode spin
  qubit},\ }\href {https://doi.org/10.1103/PhysRevB.100.125430} {\bibfield
  {journal} {\bibinfo  {journal} {Physical Review B}\ }\textbf {\bibinfo
  {volume} {100}},\ \bibinfo {pages} {125430} (\bibinfo {year} {2019})},\
  \bibinfo {note} {publisher: American Physical Society}\BibitemShut {NoStop}%
\bibitem [{\citenamefont {Hu}\ and\ \citenamefont
  {Das~Sarma}(2001)}]{hu_spin-based_2001}%
  \BibitemOpen
  \bibfield  {author} {\bibinfo {author} {\bibfnamefont {X.}~\bibnamefont
  {Hu}}\ and\ \bibinfo {author} {\bibfnamefont {S.}~\bibnamefont {Das~Sarma}},\
  }\bibfield  {title} {\bibinfo {title} {Spin-based quantum computation in
  multielectron quantum dots},\ }\href
  {https://doi.org/10.1103/PhysRevA.64.042312} {\bibfield  {journal} {\bibinfo
  {journal} {Physical Review A}\ }\textbf {\bibinfo {volume} {64}},\ \bibinfo
  {pages} {042312} (\bibinfo {year} {2001})},\ \bibinfo {note} {publisher:
  American Physical Society}\BibitemShut {NoStop}%
\bibitem [{\citenamefont {Nielsen}\ \emph {et~al.}(2013)\citenamefont
  {Nielsen}, \citenamefont {Barnes}, \citenamefont {Kestner},\ and\
  \citenamefont {Das~Sarma}}]{nielsen_six-electron_2013}%
  \BibitemOpen
  \bibfield  {author} {\bibinfo {author} {\bibfnamefont {E.}~\bibnamefont
  {Nielsen}}, \bibinfo {author} {\bibfnamefont {E.}~\bibnamefont {Barnes}},
  \bibinfo {author} {\bibfnamefont {J.~P.}\ \bibnamefont {Kestner}},\ and\
  \bibinfo {author} {\bibfnamefont {S.}~\bibnamefont {Das~Sarma}},\ }\bibfield
  {title} {\bibinfo {title} {Six-electron semiconductor double quantum dot
  qubits},\ }\href {https://doi.org/10.1103/PhysRevB.88.195131} {\bibfield
  {journal} {\bibinfo  {journal} {Physical Review B}\ }\textbf {\bibinfo
  {volume} {88}},\ \bibinfo {pages} {195131} (\bibinfo {year} {2013})},\
  \bibinfo {note} {publisher: American Physical Society}\BibitemShut {NoStop}%
\bibitem [{\citenamefont {Zajac}\ \emph {et~al.}(2018)\citenamefont {Zajac},
  \citenamefont {Sigillito}, \citenamefont {Russ}, \citenamefont {Borjans},
  \citenamefont {Taylor}, \citenamefont {Burkard},\ and\ \citenamefont
  {Petta}}]{zajac_resonantly_2018}%
  \BibitemOpen
  \bibfield  {author} {\bibinfo {author} {\bibfnamefont {D.~M.}\ \bibnamefont
  {Zajac}}, \bibinfo {author} {\bibfnamefont {A.~J.}\ \bibnamefont
  {Sigillito}}, \bibinfo {author} {\bibfnamefont {M.}~\bibnamefont {Russ}},
  \bibinfo {author} {\bibfnamefont {F.}~\bibnamefont {Borjans}}, \bibinfo
  {author} {\bibfnamefont {J.~M.}\ \bibnamefont {Taylor}}, \bibinfo {author}
  {\bibfnamefont {G.}~\bibnamefont {Burkard}},\ and\ \bibinfo {author}
  {\bibfnamefont {J.~R.}\ \bibnamefont {Petta}},\ }\bibfield  {title} {\bibinfo
  {title} {Resonantly driven {CNOT} gate for electron spins},\ }\href
  {https://doi.org/10.1126/science.aao5965} {\bibfield  {journal} {\bibinfo
  {journal} {Science}\ }\textbf {\bibinfo {volume} {359}},\ \bibinfo {pages}
  {439} (\bibinfo {year} {2018})},\ \bibinfo {note} {publisher: American
  Association for the Advancement of Science}\BibitemShut {NoStop}%
\bibitem [{\citenamefont {Taylor}\ and\ \citenamefont
  {Das~Sarma}(2024)}]{taylor_neural_2024}%
  \BibitemOpen
  \bibfield  {author} {\bibinfo {author} {\bibfnamefont {J.~R.}\ \bibnamefont
  {Taylor}}\ and\ \bibinfo {author} {\bibfnamefont {S.}~\bibnamefont
  {Das~Sarma}},\ }\href {https://doi.org/10.48550/arXiv.2405.04524} {\bibinfo
  {title} {Neural network based deep learning analysis of semiconductor quantum
  dot qubits for automated control}} (\bibinfo {year} {2024}),\ \bibinfo {note}
  {arXiv:2405.04524 [cond-mat, physics:quant-ph]}\BibitemShut {NoStop}%
\bibitem [{\citenamefont {Ziegler}\ \emph {et~al.}(2023)\citenamefont
  {Ziegler}, \citenamefont {Luthi}, \citenamefont {Ramsey}, \citenamefont
  {Borjans}, \citenamefont {Zheng},\ and\ \citenamefont
  {Zwolak}}]{ziegler_tuning_2023}%
  \BibitemOpen
  \bibfield  {author} {\bibinfo {author} {\bibfnamefont {J.}~\bibnamefont
  {Ziegler}}, \bibinfo {author} {\bibfnamefont {F.}~\bibnamefont {Luthi}},
  \bibinfo {author} {\bibfnamefont {M.}~\bibnamefont {Ramsey}}, \bibinfo
  {author} {\bibfnamefont {F.}~\bibnamefont {Borjans}}, \bibinfo {author}
  {\bibfnamefont {G.}~\bibnamefont {Zheng}},\ and\ \bibinfo {author}
  {\bibfnamefont {J.~P.}\ \bibnamefont {Zwolak}},\ }\bibfield  {title}
  {\bibinfo {title} {Tuning {Arrays} with {Rays}: {Physics}-{Informed} {Tuning}
  of {Quantum} {Dot} {Charge} {States}},\ }\href
  {https://doi.org/10.1103/PhysRevApplied.20.034067} {\bibfield  {journal}
  {\bibinfo  {journal} {Physical Review Applied}\ }\textbf {\bibinfo {volume}
  {20}},\ \bibinfo {pages} {034067} (\bibinfo {year} {2023})},\ \bibinfo {note}
  {publisher: American Physical Society}\BibitemShut {NoStop}%
\bibitem [{\citenamefont {Zwolak}\ and\ \citenamefont
  {Taylor}(2023)}]{zwolak_colloquium_2023}%
  \BibitemOpen
  \bibfield  {author} {\bibinfo {author} {\bibfnamefont {J.~P.}\ \bibnamefont
  {Zwolak}}\ and\ \bibinfo {author} {\bibfnamefont {J.~M.}\ \bibnamefont
  {Taylor}},\ }\bibfield  {title} {\bibinfo {title} {\textit{{Colloquium}} :
  {Advances} in automation of quantum dot devices control},\ }\href
  {https://doi.org/10.1103/RevModPhys.95.011006} {\bibfield  {journal}
  {\bibinfo  {journal} {Reviews of Modern Physics}\ }\textbf {\bibinfo {volume}
  {95}},\ \bibinfo {pages} {011006} (\bibinfo {year} {2023})}\BibitemShut
  {NoStop}%
\bibitem [{\citenamefont {Yang}\ \emph {et~al.}(2011)\citenamefont {Yang},
  \citenamefont {Wang},\ and\ \citenamefont {Das~Sarma}}]{yang_generic_2011}%
  \BibitemOpen
  \bibfield  {author} {\bibinfo {author} {\bibfnamefont {S.}~\bibnamefont
  {Yang}}, \bibinfo {author} {\bibfnamefont {X.}~\bibnamefont {Wang}},\ and\
  \bibinfo {author} {\bibfnamefont {S.}~\bibnamefont {Das~Sarma}},\ }\bibfield
  {title} {\bibinfo {title} {Generic {Hubbard} model description of
  semiconductor quantum dot spin qubits},\ }\href
  {https://doi.org/10.1103/PhysRevB.83.161301} {\bibfield  {journal} {\bibinfo
  {journal} {Physical Review B}\ }\textbf {\bibinfo {volume} {83}},\ \bibinfo
  {pages} {161301(R)} (\bibinfo {year} {2011})}\BibitemShut {NoStop}%
\bibitem [{\citenamefont {Wang}\ \emph {et~al.}(2011)\citenamefont {Wang},
  \citenamefont {Yang},\ and\ \citenamefont {Das~Sarma}}]{wang_quantum_2011}%
  \BibitemOpen
  \bibfield  {author} {\bibinfo {author} {\bibfnamefont {X.}~\bibnamefont
  {Wang}}, \bibinfo {author} {\bibfnamefont {S.}~\bibnamefont {Yang}},\ and\
  \bibinfo {author} {\bibfnamefont {S.}~\bibnamefont {Das~Sarma}},\ }\bibfield
  {title} {\bibinfo {title} {Quantum theory of the charge stability diagram of
  semiconductor double quantum dot systems},\ }\href@noop {} {\bibfield
  {journal} {\bibinfo  {journal} {Physical Review B}\ }\textbf {\bibinfo
  {volume} {84}},\ \bibinfo {pages} {115301} (\bibinfo {year}
  {2011})}\BibitemShut {NoStop}%
\bibitem [{\citenamefont {Das~Sarma}\ \emph {et~al.}(2011)\citenamefont
  {Das~Sarma}, \citenamefont {Wang},\ and\ \citenamefont
  {Yang}}]{das_sarma_hubbard_2011}%
  \BibitemOpen
  \bibfield  {author} {\bibinfo {author} {\bibfnamefont {S.}~\bibnamefont
  {Das~Sarma}}, \bibinfo {author} {\bibfnamefont {X.}~\bibnamefont {Wang}},\
  and\ \bibinfo {author} {\bibfnamefont {S.}~\bibnamefont {Yang}},\ }\bibfield
  {title} {\bibinfo {title} {Hubbard model description of silicon spin qubits:
  charge stability diagram and tunnel coupling in {Si} double quantum dots},\
  }\href@noop {} {\bibfield  {journal} {\bibinfo  {journal} {Physical Review
  B}\ }\textbf {\bibinfo {volume} {83}},\ \bibinfo {pages} {235314} (\bibinfo
  {year} {2011})}\BibitemShut {NoStop}%
\bibitem [{\citenamefont {Zwolak}\ \emph {et~al.}(2020)\citenamefont {Zwolak},
  \citenamefont {McJunkin}, \citenamefont {Kalantre}, \citenamefont {Dodson},
  \citenamefont {MacQuarrie}, \citenamefont {Savage}, \citenamefont {Lagally},
  \citenamefont {Coppersmith}, \citenamefont {Eriksson},\ and\ \citenamefont
  {Taylor}}]{zwolak_autotuning_2020}%
  \BibitemOpen
  \bibfield  {author} {\bibinfo {author} {\bibfnamefont {J.~P.}\ \bibnamefont
  {Zwolak}}, \bibinfo {author} {\bibfnamefont {T.}~\bibnamefont {McJunkin}},
  \bibinfo {author} {\bibfnamefont {S.~S.}\ \bibnamefont {Kalantre}}, \bibinfo
  {author} {\bibfnamefont {J.~P.}\ \bibnamefont {Dodson}}, \bibinfo {author}
  {\bibfnamefont {E.~R.}\ \bibnamefont {MacQuarrie}}, \bibinfo {author}
  {\bibfnamefont {D.~E.}\ \bibnamefont {Savage}}, \bibinfo {author}
  {\bibfnamefont {M.~G.}\ \bibnamefont {Lagally}}, \bibinfo {author}
  {\bibfnamefont {S.~N.}\ \bibnamefont {Coppersmith}}, \bibinfo {author}
  {\bibfnamefont {M.~A.}\ \bibnamefont {Eriksson}},\ and\ \bibinfo {author}
  {\bibfnamefont {J.~M.}\ \bibnamefont {Taylor}},\ }\bibfield  {title}
  {\bibinfo {title} {Autotuning of double-dot devices in situ with machine
  learning},\ }\href {https://doi.org/10.1103/PhysRevApplied.13.034075}
  {\bibfield  {journal} {\bibinfo  {journal} {Phys. Rev. Appl.}\ }\textbf
  {\bibinfo {volume} {13}},\ \bibinfo {pages} {034075} (\bibinfo {year}
  {2020})}\BibitemShut {NoStop}%
\bibitem [{\citenamefont {Barnes}\ \emph {et~al.}(2011)\citenamefont {Barnes},
  \citenamefont {Kestner}, \citenamefont {Nguyen},\ and\ \citenamefont
  {Das~Sarma}}]{barnes_screening_2011}%
  \BibitemOpen
  \bibfield  {author} {\bibinfo {author} {\bibfnamefont {E.}~\bibnamefont
  {Barnes}}, \bibinfo {author} {\bibfnamefont {J.~P.}\ \bibnamefont {Kestner}},
  \bibinfo {author} {\bibfnamefont {N.~T.~T.}\ \bibnamefont {Nguyen}},\ and\
  \bibinfo {author} {\bibfnamefont {S.}~\bibnamefont {Das~Sarma}},\ }\bibfield
  {title} {\bibinfo {title} {Screening of charged impurities with multielectron
  singlet-triplet spin qubits in quantum dots},\ }\href
  {https://doi.org/10.1103/PhysRevB.84.235309} {\bibfield  {journal} {\bibinfo
  {journal} {Physical Review B}\ }\textbf {\bibinfo {volume} {84}},\ \bibinfo
  {pages} {235309} (\bibinfo {year} {2011})}\BibitemShut {NoStop}%
\bibitem [{\citenamefont {Szabó}\ and\ \citenamefont
  {Ostlund}(1982)}]{szabo_modern_1982}%
  \BibitemOpen
  \bibfield  {author} {\bibinfo {author} {\bibfnamefont {A.}~\bibnamefont
  {Szabó}}\ and\ \bibinfo {author} {\bibfnamefont {N.~S.}\ \bibnamefont
  {Ostlund}},\ }\href@noop {} {\emph {\bibinfo {title} {Modern {Quantum}
  {Chemistry}: {Introduction} to {Advanced} {Electronic} {Structure}
  {Theory}}}}\ (\bibinfo  {publisher} {Macmillan},\ \bibinfo {year}
  {1982})\BibitemShut {NoStop}%
\bibitem [{\citenamefont {Davies}\ \emph {et~al.}(1995)\citenamefont {Davies},
  \citenamefont {Larkin},\ and\ \citenamefont
  {Sukhorukov}}]{davies_modeling_1995}%
  \BibitemOpen
  \bibfield  {author} {\bibinfo {author} {\bibfnamefont {J.~H.}\ \bibnamefont
  {Davies}}, \bibinfo {author} {\bibfnamefont {I.~A.}\ \bibnamefont {Larkin}},\
  and\ \bibinfo {author} {\bibfnamefont {E.~V.}\ \bibnamefont {Sukhorukov}},\
  }\bibfield  {title} {\bibinfo {title} {Modeling the patterned two-dimensional
  electron gas: {Electrostatics}},\ }\href {https://doi.org/10.1063/1.359446}
  {\bibfield  {journal} {\bibinfo  {journal} {Journal of Applied Physics}\
  }\textbf {\bibinfo {volume} {77}},\ \bibinfo {pages} {4504} (\bibinfo {year}
  {1995})}\BibitemShut {NoStop}%
\bibitem [{\citenamefont {Anderson}\ \emph {et~al.}(2022)\citenamefont
  {Anderson}, \citenamefont {Gyure}, \citenamefont {Quinn}, \citenamefont
  {Pan}, \citenamefont {Ross},\ and\ \citenamefont
  {Kiselev}}]{anderson_high-precision_2022}%
  \BibitemOpen
  \bibfield  {author} {\bibinfo {author} {\bibfnamefont {C.~R.}\ \bibnamefont
  {Anderson}}, \bibinfo {author} {\bibfnamefont {M.~F.}\ \bibnamefont {Gyure}},
  \bibinfo {author} {\bibfnamefont {S.}~\bibnamefont {Quinn}}, \bibinfo
  {author} {\bibfnamefont {A.}~\bibnamefont {Pan}}, \bibinfo {author}
  {\bibfnamefont {R.~S.}\ \bibnamefont {Ross}},\ and\ \bibinfo {author}
  {\bibfnamefont {A.~A.}\ \bibnamefont {Kiselev}},\ }\bibfield  {title}
  {\bibinfo {title} {High-precision real-space simulation of
  electrostatically-confined few-electron states},\ }\href@noop {} {\bibfield
  {journal} {\bibinfo  {journal} {AIP Advances}\ }\textbf {\bibinfo {volume}
  {12}},\ \bibinfo {pages} {065123} (\bibinfo {year} {2022})}\BibitemShut
  {NoStop}%
\bibitem [{\citenamefont {Nielsen}\ and\ \citenamefont
  {Chuang}(2010)}]{nielsen_quantum_2010}%
  \BibitemOpen
  \bibfield  {author} {\bibinfo {author} {\bibfnamefont {M.~A.}\ \bibnamefont
  {Nielsen}}\ and\ \bibinfo {author} {\bibfnamefont {I.~L.}\ \bibnamefont
  {Chuang}},\ }\href@noop {} {\emph {\bibinfo {title} {Quantum computation and
  quantum information}}}\ (\bibinfo  {publisher} {Cambridge university press},\
  \bibinfo {year} {2010})\BibitemShut {NoStop}%
\bibitem [{\citenamefont {Zwolak}\ \emph {et~al.}(2018)\citenamefont {Zwolak},
  \citenamefont {Kalantre}, \citenamefont {Wu}, \citenamefont {Ragole},\ and\
  \citenamefont {Taylor}}]{zwolak_qflow_2018}%
  \BibitemOpen
  \bibfield  {author} {\bibinfo {author} {\bibfnamefont {J.~P.}\ \bibnamefont
  {Zwolak}}, \bibinfo {author} {\bibfnamefont {S.~S.}\ \bibnamefont
  {Kalantre}}, \bibinfo {author} {\bibfnamefont {X.}~\bibnamefont {Wu}},
  \bibinfo {author} {\bibfnamefont {S.}~\bibnamefont {Ragole}},\ and\ \bibinfo
  {author} {\bibfnamefont {J.~M.}\ \bibnamefont {Taylor}},\ }\bibfield  {title}
  {\bibinfo {title} {{QFlow} lite dataset: {A} machine-learning approach to the
  charge states in quantum dot experiments},\ }\href@noop {} {\bibfield
  {journal} {\bibinfo  {journal} {PLOS ONE}\ }\textbf {\bibinfo {volume}
  {13}},\ \bibinfo {pages} {e0205844} (\bibinfo {year} {2018})}\BibitemShut
  {NoStop}%
\end{thebibliography}%
\end{document}